\documentclass
[preprintnumbers,reprint,twocolumn,prd,amsmath,amssymb,nofootinbib]
{revtex4-1}

\usepackage{graphicx}
\usepackage[tight]{subfigure}

\newcommand{\incgraph}[2][]{%
  \IfFileExists{pr/#2.pdf}{%
    \includegraphics[#1]{pr/#2.pdf}}{%
    \includegraphics[#1]{#2.pdf}}}

\newcommand{\re}{\mathrm{Re}}
\newcommand{\wt}{\widetilde}

\newcommand{\TeV}{\ \mathrm{TeV}}
\newcommand{\GeV}{\ \mathrm{GeV}}

\newcommand{\tb}{\tan\!\beta}

\newcommand{\deltax}[1]{\delta^{#1}}
\newcommand{\deltau}{\deltax{u}}
\newcommand{\deltad}{\deltax{d}}
\newcommand{\deltae}{\deltax{e}}
\newcommand{\deltaf}{\deltax{f}}
\newcommand{\deu}[2]{(\deltau_{#1})_{#2}}
\newcommand{\ded}[2]{(\deltad_{#1})_{#2}}
\newcommand{\dee}[2]{(\deltae_{#1})_{#2}}
\newcommand{\defermion}[2]{(\deltaf_{#1})_{#2}}
\newcommand{\Ax}[1]{A^{#1}}
\newcommand{\Aup}{\Ax{u}}
\newcommand{\Adown}{\Ax{d}}
\newcommand{\Alepton}{\Ax{e}}
\newcommand{\Afermion}{\Ax{f}}
\newcommand{\Adownlepton}{\Ax{e,d}}
\newcommand{\Au}[1]{\Aup_{#1}}
\newcommand{\Ad}[1]{\Adown_{#1}}
\newcommand{\Ae}[1]{\Alepton_{#1}}
\newcommand{\Af}[1]{\Afermion_{#1}}
\newcommand{\Ade}[1]{\Adownlepton_{#1}}
\newcommand{\msl}{m_{\widetilde{l}}}
\newcommand{\msq}{m_{\widetilde{q}}}
\newcommand{\msf}{m_{\widetilde{f}}}
\newcommand{\mgl}{M_3}

\newcommand{\MH}{m_{H^0}}
\newcommand{\SE}{S}
\newcommand{\DSE}{\Delta\SE}
\newcommand{\onel}{\mathrm{1L}}
\newcommand{\SEonel}{S_\onel}
\newcommand{\Vonel}{V_\onel}
\newcommand{\phibounce}{\overline{\phi}}

\newcommand{\phie}{\phi_e}
\newcommand{\phieinit}{\phie}
\newcommand{\phifv}{\phi_+}
\newcommand{\cosmoconst}{\Omega}

\newcommand{\BR}{B}
\newcommand{\tmg}{\tau \rightarrow \mu \gamma}
\newcommand{\teg}{\tau \rightarrow e \gamma}
\newcommand{\meg}{\mu \rightarrow e \gamma}
\newcommand{\bsg}{B \rightarrow X_s \gamma}

\newcommand{\DRbarprime}{$\overline{\text{DR}}'$}

\newlength{\squarefiguresize}
\newlength{\rectangularfigureheight}
\begin{document}
%\preprint{arXiv:1011.????}
\preprint{DESY 10--214}
\title{Metastability bounds on flavour-violating trilinear soft terms
  in the MSSM}
\author{Jae-hyeon~Park}
\affiliation{Deutsches Elektronen-Synchrotron DESY,
  Notkestra{\ss}e 85, 22607 Hamburg, Germany}
\begin{abstract}
The vacuum stability bounds on flavour-violating trilinear soft terms
are revisited from the viewpoint that one should not ban
a standard-model-like false vacuum
as long as it is long-lived on a cosmological timescale.
The vacuum transition rate is evaluated numerically by searching for
the bounce configuration.
Like stability, a metastability bound does not decouple
even if sfermion masses grow.
Apart from being more generous than stability,
the new bounds are largely independent of
Yukawa couplings except for the stop trilinears.
With vacuum longevity imposed on otherwise arbitrary $LR$ insertions,
it is found that a super flavour factory has the potential to probe
sparticle masses up to a few TeV through $B$ and $\tau$ physics whereas
the MEG experiment might cover a far wider range.
In the stop sector, metastability is more restrictive than
any existing experimental constraint such as from
electroweak precision data.
Also discussed are dependency on other parameters and
reliability under radiative corrections.
%  Prospect of future indirect searches for supersymmetric particles
%  is essentially determined by sizes of their mixings,
%  as these are the sources of new flavor violating effects---%
%  the larger the mixings are,
%  the heavier sparticles come within the reach of a super flavor factory.
\end{abstract}
\maketitle

\section{Introduction}

% potential of flavor physics probing new physics
%% simple minded estimation in the presence of large LR
% super B factory and its CDR
%% Plots of MI vs gluino mass
% scaling of FV processes as a function of LL/RR, LR/RL insertions
%% LL/RR decouples faster than LR/RL
%% -> LR/RL more important for heavy sparticles
% scaling of bounds from FCNC
%% (the only) model independent theoretical bound: vacuum stability
%% FCNC bounds decouple but vacuum bounds don't
%% -> vacuum bounds as input for prediction of FCNC
% what to do with vacuum, stability or metastability?
%% -> metastability is more sensible
%% inflationary effect supports this view
% History of studies on (meta)stability in MSSM
%% Traditional bound, FCNC bound, metastability bound
%% missing piece is FV meta: subject of this work
% organization of paper

Flavour physics is an important means to test any theoretical description
of elementary particles.
Among numerous proposals, the one still most compelling would be
% arguably
the Minimal Supersymmetric Standard Model (MSSM) that
offers solutions to problems not addressed by the Standard Model (SM)\@.
In the MSSM, the Lagrangian has soft supersymmetry breaking terms.
Among them, the scalar mass and the trilinear terms are
two representative sources of flavour/$CP$ violation apart from
the Cabibbo-Kobayashi-Maskawa (CKM) matrix.
Unless tuned in a particular way, they could add to
an existing flavour-changing-neutral-current (FCNC) process
such as $\bsg$ or create a new one, e.g.\ $\meg$,
through loop corrections.
In this way, flavour physics provides an invaluable input
to the construction of an underlying theory that
implements spontaneous super- or flavour symmetry breaking.

For the same reason,
flavour physics has been a traditional tool to unveil supersymmetry.
Now that the Large Hadron Collider (LHC) has started its successful
operation, the significance of this role has diminished.
Nevertheless, there are at least two areas in which flavour studies
could make contributions.
The first is to scrutinise the flavour structure of newly discovered particles.
This is a complement to the LHC that is less sensitive to
extra flavour/$CP$ violation.
The second is to continue the search for supersymmetric particles.
This should be vital particularly in a scenario where sparticle masses are
around the LHC reach or beyond.
% In either case, the immediate question is how probable it is to
% observe a discrepancy in an FCNC process for a given sparticle mass.
% It depends on the size of a flavour-violating term in the Lagrangian,
% often represented by the dimensionless quantity $\delta$
% to be defined in the next section.
In either case, the crucial quantity is the size of
a flavour-violating term in the Lagrangian,
often represented by the dimensionless quantity $\delta$
to be defined in the next section.

According to the chirality structure, a mass insertion (MI)
can be classified as either chirality-preserving or chirality-flipping.
In a high sparticle mass region,
the latter type becomes generically more important than the former
\cite{Kane et al}.
Taking for example the ratio of the one-loop $\bsg$ amplitudes caused by
the $LR$ and the $LL$ insertions, one finds a quantity proportional to
$(\mgl/m_b) [\ded{23}{LR}/\ded{23}{LL}]$
(see e.g.\ Ref.~\cite{Gabbiani:1996hi}).
It shows that the relative contribution from the $LR$ insertion
grows linearly as the gluino mass $\mgl$.
This is one of the reasons why the subject of this study is
the trilinear terms.

The obvious problem is that
one needs to get an idea of the flavour structure of particles
that might yet be found.
A popular strategy is to employ a model or ansatz that leads to
a particular pattern of the soft terms.
There are many works in this approach,
and this article is not going to be another.
The plan here is the following.
The most optimistic scenario is presumed in which
any $LR$ insertion can take an arbitrary value
as long as it stays within existing experimental and theoretical limits.
Using this input,
one can estimate the maximum possible new signal in an FCNC process.
Then it is compared with sensitivities of running or forthcoming
experimental searches.

As for the input mentioned above,
one would learn nothing new if only the present FCNC limits were used.
One needs information from outside flavour physics.
There are such theoretical bounds that arise from vacuum stability.
If a trilinear term is too large,
the MSSM scalar potential develops
a charge-and/or-colour-breaking (CCB) vacuum
deeper than the the standard-model-like (SML) local minimum
\cite{ccb,Casas:1996de}
or an unbounded-from-below (UFB) direction in the field space
\cite{Casas:1996de}.
Most notably, the conceptual design report of a super flavour factory (SFF)
has made an assessment of its ability to reconstruct
a $LR$ squark insertion, imposing these bounds
\cite{Bona:2007qt}.

However, it is questionable whether the vacuum stability is an essential
requirement in particular when one is performing an analysis
of generic soft terms that is supposed to be objective.
More sensibly,
one could use the lifetime of the SML vacuum as the criterion.
This viewpoint would not make sense if the universe were driven
away from there in its history.
In the case of a squark direction, its temperature-dependent mass
\cite{Quiros:1994dr}
may defer the formation of a CCB minimum until the electroweak phase transition
has happened \cite{Kusenko:1996jn,thermal mass}.
Supersymmetry breaking by inflation provides
a more general mechanism to lift flat directions so that
relaxation to the symmetric phase can take place
\cite{inflationary mass}.
The bottom line is that there seems to be no definite reason to preclude
a lower energy CCB state just for its existence.
% , apart from psychological effects.
In this approach,
there have been attempts to obtain a viable parameter space of
the flavour-conserving stop trilinear \cite{Claudson:1983et,Kusenko:1996jn}
or the stau trilinear coupling \cite{A stau},
the sparticle \cite{Riotto:1995am,Strumia:1996pr} or
the Higgs soft masses \cite{Evans:2008zx},
or messenger couplings for gauge mediation \cite{gauge mediation}.
This work is an application of the same idea to
the flavour-violating trilinear couplings.

The rest of the article begins with an abridged account of
theoretical backgrounds in section~\ref{sec:theory}.
In section~\ref{sec:analysis}, more practical details of
the analysis method are given.
The metastability bounds thus obtained are presented
in section~\ref{sec:results} in which their behaviours are also explained.
Section~\ref{sec:comparison} is devoted to
an application of the findings to flavour physics.
Finally, the article is summarised in section~\ref{sec:conc}.

%\section{Theoretical background}
\section{Theory}
\label{sec:theory}

For concise presentation, % of $D$-flat directions,
the following notations shall be employed.
The species of a given matter field is denoted by $f$
which can be one of $e,d,u$, standing for
charged leptons, down-, and up-type quarks, respectively.
A field symbol without the superscript $c$ refers to a component of
the SU(2) doublets,
$L = (\nu, e), Q = (u, d), H_d = (H_d^0, H_d^-), H_u = (H_u^+, H_u^0)$.
Also, another redundant symbol for the down-type Higgs doublet is introduced,
i.e.\ $H_e \equiv H_d$.

% CCB/UFB directions in MSSM
% FV CCB/UFB bounds
%% argue: UFB direction irrelevant
% semiclassical approximation of tunneling rate
%% criterion: B < 400
%% position of true vacuum does not matter
% radiative corrections
%% suppressed by choice of scale
%%% relevant field values not far from TeV
%% schemes DR bar, *DR bar prime*

The scalar potential of the MSSM is in the form
\begin{align}
  \label{eq:VMSSM}
  V &= V_D + V_F + V_\mathrm{soft} ,
\intertext{where the $D$-, the $F$-, and the soft term contributions read}
  \label{eq:VD}
  V_D &= \frac{1}{2} \sum_a g_a^2 \Bigl(
  \sum_\alpha \phi_\alpha^\dagger T^a \phi_\alpha
  \Bigr)^2 ,
  \\
  \label{eq:VF}
  V_F &= \sum_\alpha
  \left| \frac{\partial W}{\partial \phi_\alpha} \right|^2 ,
  \\
  V_\mathrm{soft} &=
  \wt{L}^\dagger_i    (M^2_L)_{ij}    \, \wt{L}_j   +
  \wt{e}^{c\dagger}_i (M^2_{e^c})_{ij}\, \wt{e}^c_j
  \nonumber
  \\ \nonumber
  &+
  \wt{Q}^\dagger_i    (M^2_Q)_{ij}    \, \wt{Q}_j   +
  \wt{d}^{c\dagger}_i (M^2_{d^c})_{ij}\, \wt{d}^c_j +
  \wt{u}^{c\dagger}_i (M^2_{u^c})_{ij}\, \wt{u}^c_j
  \\ \nonumber
  &+ 2\,\re \bigl[
    H_d\, \wt{L}_i \Ae{ij}\, \wt{e}^c_j
  + H_d\, \wt{Q}_i \Ad{ij}\, \wt{d}^c_j
  - H_u\, \wt{Q}_i \Au{ij}\, \wt{u}^c_j
%  \bigr]
%  \\
%  &+ 2\,\re \bigl[
  \bigr]
  \\
  &+ m^2_{H_d} |H_d|^2 + m^2_{H_u} |H_u|^2 + 2\,\re \bigl[b \, H_u H_d\bigr] ,
  \\
\intertext{with the superpotential}
\begin{split}
  W &=
    H_d\, L_i \lambda^e_{ij}\, e^c_j
  + H_d\, Q_i \lambda^d_{ij}\, d^c_j
  - H_u\, Q_i \lambda^u_{ij}\, u^c_j \\
 &+ \mu\, H_u H_d .
\end{split}
\end{align}
The index $\alpha$ in~(\ref{eq:VD}) and (\ref{eq:VF})
runs over each superfield in the model and $\phi_\alpha$ is
its scalar component.

For discussion of supersymmetric flavour violation,
it is convenient to use the language of mass insertion approximation
\cite{Hall:1985dx}.
In order to specify the families of a given species $f$, one should take
% The families are with respect to
the super-CKM basis that leads to
the diagonal Yukawa matrix, $\lambda^f_{ij} = \lambda_{f_i} \delta_{ij}$
with positive eigenvalues.
In this article,
a $LR$ mass insertion is related with a trilinear coupling by % the formula
\begin{align}
  \label{eq:delta}
  \defermion{ij}{RL} &= \frac{\Af{ji} \langle H_f^0 \rangle}
%  {\msf^2} ,
%  {(M^2_f)_{jj} + (M^2_{f^c})_{ii}},
  {(M^2_\mathrm{av})^f_{ji}},
  \\
(M^2_\mathrm{av})^f_{ij} &\equiv \frac{1}{2}
\bigl[(M^2_f)_{ii} + (M^2_{f^c})_{jj}\bigr] .
\end{align}
% where $(M^2_\mathrm{av})^f_{ij} \equiv [(M^2_f)_{ii} + (M^2_{f^c})_{jj}]/2$.
Note that this definition of average mass is different from
that in Ref.~\cite{Gabbiani:1996hi}, i.e.\ the geometric mean.
Nevertheless, this variance does not matter since
the soft scalar masses shall be assumed to be nearly degenerate.
A trivial but useful identity obeyed by the $LR$ insertions is
$\defermion{ij}{LR} = \defermion{ji}{RL}^*$, which follows from the
Hermiticity of a sfermion mass matrix.

It should be instructive to review
the stability bounds on flavour-violating trilinear couplings
\cite{Casas:1996de},
as the potential along a CCB direction helps
to understand qualitatively
many of the properties of the tunnelling bounds.
Suppose that the scalar fields take on values such that
% $|H_d^0| = |\wt{e}_2| = |\wt{e}^c_3| = a$
% $|H_d^0| = |\wt{d}_2| = |\wt{d}^c_3| = a$
% $|H_u^0| = |\wt{u}_2| = |\wt{u}^c_3| = a$
\begin{equation}
  \label{eq:D-flat direction}
  \bigl|H_f^0\bigr| = \bigl|\wt{f}_i\bigr| = \bigl|\wt{f}^c_j\bigr| = a ,
\end{equation}
for one particular set of $f$, $i$, and $j$, where
$i,j = 1,2,3,$ are family indices
% The families are 
with respect to the super-CKM basis.  % that leads to
% the diagonal Yukawa matrix, $\lambda^f_{ij} = \lambda_{f_i} \delta_{ij}$
% with positive eigenvalues.
The other fields are all assumed to vanish.
Then one can check that $V_D = 0$ and
the lowest potential energy subject to the above condition is
\begin{equation}
  \label{eq:D-flat potential}
  \begin{aligned}
  V_\mathrm{L.E.} &=
%  \bigl[ (M^2_f)_{ii} + (M^2_{f^c})_{jj} + m^2_{H_f} + |\mu|^2 \bigr]\, a^2
  \bigl[ 2\,(M^2_\mathrm{av})^f_{ij} + m^2_{H_f} + |\mu|^2 \bigr]\, a^2
  \\
  &- 2 \bigl|\Af{ij}\bigr|\, a^3
  + \bigl(\lambda_{f_i}^2 + \lambda_{f_j}^2\bigr)\, a^4 .
  \end{aligned}
  % \\
  % V_\mathrm{L.E.} =
  % [ (M^2_L)_2 + (M^2_{e^c})_3 + m^2_{H_d} + |\mu|^2]\, a^2
  % - 2 |\Ae{23}|\, a^3
  % + (\lambda_\mu^2 + \lambda_\tau^2)\, a^4
  % \\
  % V_\mathrm{L.E.} =
  % [ (M^2_Q)_2 + (M^2_{d^c})_3 + m^2_{H_d} + |\mu|^2]\, a^2
  % - 2 |\Ad{23}|\, a^3
  % + (\lambda_s^2 + \lambda_b^2)\, a^4
  % \\
  % V_\mathrm{L.E.} =
  % [ (M^2_Q)_2 + (M^2_{u^c})_3 + m^2_{H_u} + |\mu|^2]\, a^2
  % - 2 |\Au{23}|\, a^3
  % + (\lambda_c^2 + \lambda_t^2)\, a^4
\end{equation}
Due to the cubic term in $a$,
a minimum can appear that is deeper than the SML vacuum
unless
% \begin{equation}
%   |\Ad{23}|^2 < \lambda_b^2
%   [(M^2_Q)_2 + (M^2_{d^c})_3 + m^2_{H_d} + |\mu|^2]
% \end{equation}
\begin{equation}
  \label{eq:stability bound on A}
\bigl|\Af{ij}\bigr|^2 < \lambda_{f_{\max(i,j)}}^2
%\bigl[(M^2_f)_{ii} + (M^2_{f^c})_{jj} + m^2_{H_f} + |\mu|^2\bigr] .
\bigl[2\,(M^2_\mathrm{av})^f_{ij} + m^2_{H_f} + |\mu|^2\bigr] .
% k = \max(i,j)
\end{equation}
%       \begin{equation}
% \begin{aligned}
% |\Au{ij}|^2&<\lambda_{u_k}^2[(M^2_Q)_i+(M^2_{u^c})_j+m^2_{H_u}+|\mu|^2],&
% k &= \max(i,j) \\
% |\Ad{ij}|^2&<\lambda_{d_k}^2[(M^2_Q)_i+(M^2_{d^c})_j+m^2_{H_d}+|\mu|^2],&
% k &= \max(i,j) \\
% |\Ae{ij}|^2&<\lambda_{e_k}^2[(M^2_L)_i+(M^2_{e^c})_j+m^2_{H_d}+|\mu|^2],&
% k &= \max(i,j)
% \end{aligned}
%       \end{equation}
By imposing this inequality, one can avoid a CCB minimum.

This theoretical bound does not decouple even if sfermions grow heavier.
On the contrary, it gets stronger.
This property is most easily demonstrated in terms of a $\delta$ parameter.
Using~(\ref{eq:delta}),
one can translate~(\ref{eq:stability bound on A}) into the form
% \begin{align}
%   \label{eq:stability bound on d}
%   \begin{aligned}
%   \bigl|\defermion{ij}{LR}\bigr| &< m_{f_k}
%   \frac{2\bigl[(M^2_f)_{ii}+(M^2_{f^c})_{jj}+m^2_{H_f}+|\mu|^2\bigr]^{1/2}}
%   {(M^2_f)_{ii} + (M^2_{f^c})_{jj}}
%   \\
%   & \sim \frac{m_{f_k}}{\msf} ,
%   \quad
%   k = \max(i,j) ,
%   \end{aligned}
% \end{align}
\begin{equation}
  \label{eq:stability bound on d}
  \bigl|\defermion{ij}{LR}\bigr| < m_{f_{\max(i,j)}}
  \frac{\bigl[2\, (M^2_\mathrm{av})^f_{ij} + m^2_{H_f}+|\mu|^2\bigr]^{1/2}}
  {(M^2_\mathrm{av})^f_{ij}} ,
%  \\ \nonumber
%  & \sim \frac{m_{f_k}}{\msf} , \quad
%  k = \max(i,j) ,
\end{equation}
where $m_{f_k}$ is the mass of fermion $f_k$.
% signifies the representative mass of the sfermions being considered.
This bound scales as the inverse power of the average sfermion mass.
This behaviour is at variance with an FCNC limit on 
$|\defermion{ij}{LR}\bigr|$ that
grows as sfermions become heavier.
Another point to notice is that the above restriction is independent of $\tb$.
As will be shown, this is not the case with a metastability bound.

% comment on UFB
In addition to the CCB minima, UFB field directions
can appear due to large trilinear couplings \cite{Casas:1996de}.
If one is worried only about the existence of a deeper point
in the field space, both types of constraints should be taken into account.
As far as a tunnelling process is concerned, however,
a UFB direction is most likely irrelevant for the following reasons.
As one traces this direction starting from the SML vacuum,
the potential barrier is generically thicker than along a CCB direction.
Moreover, there is a section of path in which
$V_D$ does not vanish thereby making the barrier much higher as well.
Therefore, a tunnelling process prefers a path close to
a CCB direction that obeys $D$-flatness all the way.

Even though expressed in a nice analytic form,
the role of a CCB condition as a phenomenological constraint is rather unclear.
Instead, the lifetime of the SML vacuum shall be required to be
long enough.
In a semiclassical approximation \cite{tunnelling},
the decay probability of a metastable vacuum
per unit time per unit volume is given by
\begin{equation}
  \label{eq:probability}
  \Gamma/V = A\,\exp(-\SE[\phibounce]) ,
\end{equation}
where $A$ is a prefactor that needs to be guessed,
$\SE$ is the Euclidean action,
and $\phibounce$ is its bounce with $O(4)$ symmetry
\cite{Coleman:1977th}.
This symmetry allows one to use a single coordinate,
i.e.\ the radial distance $\rho$,
with respect to which the action can be written
\begin{equation}
  \label{eq:SE}
  \SE[\phi(\rho)] = 2\pi^2 \int_0^\infty d\rho\,\rho^3
  \biggl[ \left|\frac{d\phi}{d\rho}\right|^2 + V(\phi) \biggr] .
\end{equation}
Note that the normalisation of the kinetic term is that of complex fields,
as $\phi$ is intended to be eventually the MSSM scalars.
The field configuration $\phibounce(\rho)$ is a stationary point of
$\SE[\phi(\rho)]$ subject to the boundary conditions,
\begin{equation}
  \label{eq:BCs}
%  \frac{d\phibounce}{d\rho} (0) = 0, \quad
%  \lim_{\rho \rightarrow \infty}\phibounce(\rho) = \phifv
%  \frac{d\phibounce}{d\rho} (0) = 0, \quad
%  \phibounce(\infty) = \phifv
%  \lim_{\rho \rightarrow \infty} \phibounce(\rho) = \phifv, \quad
%  \left.\frac{d\phibounce}{d\rho}\right|_{\rho=0} = 0,
  \phibounce(\rho \rightarrow \infty) = \phifv, \quad
  \frac{d\phibounce}{d\rho} (\rho=0) = 0 ,
\end{equation}
where $\phifv$ is the false vacuum.

In the context of the MSSM,
$\phi$ represents the vector of scalar fields in the model,
and $\phifv$ is the metastable SML vacuum.
One can evaluate the action by replacing $V$ in~(\ref{eq:SE})
with the scalar potential~(\ref{eq:VMSSM}) plus a constant term
such that $V(\phifv) = 0$.

On the other hand, the prefactor $A$ in~(\ref{eq:probability})
is practically impossible to compute.
Based on dimensional grounds, it is guesstimated to be
the fourth power of the mass scale appearing in the problem.
In this work, $A = (100\GeV)^4$ shall be used with which
one can estimate the smallest acceptable action.
The lifetime of the whole observable universe staying at the SML vacuum
is roughly $V/(\Gamma t_0^3)$ where $t_0 \approx 10\ \mathrm{Gyr}$
is the age of the universe.
One can keep this longer than $t_0$ by demanding that
\cite{Claudson:1983et}
\begin{equation}
  \label{eq:limit on S}
  \SE > 400 .
\end{equation}

The above value of $A$ may be regarded
a little small in view of the sfermion masses in the TeV range.
Choosing a small $A$ implies a conservative
parameter space exclusion.
Nonetheless, the constraint is much less sensitive to
the variation of $A$ than of $S$.
If desired, one could translate the multiplicative uncertainty of $A$
into the additive one of $S$ by taking the logarithm.
For instance,
taking $A$ instead to be the fourth power of the reduced
Planck scale would push up the lower bound on $\SE$ by 150.

Another type of uncertainty arises from radiative corrections.
It is important to estimate their effects especially because
the numerical analysis uses the tree-level potential.
One can implement the one-loop correction to the tunnelling rate
simply by using the one-loop effective potential
\begin{equation}
  \label{eq:V1L}
  \Vonel = V + \frac{1}{16 \pi^2} V^{(1)} + \cosmoconst ,
\end{equation}
instead of the tree-level potential $V$ in~(\ref{eq:SE}).
The constant $\cosmoconst$ should be adjusted so that
% both $V$ and 
$\Vonel(\phifv) = 0$.  % vanishes at the false vacuum.
The one-loop correction in the \DRbarprime\ scheme is
\cite{V1 in DRbarprime}
\begin{equation}
  V^{(1)} = \sum_n (-1)^{2 s_n} (2 s_n + 1) \,
  \frac{m^4_n}{4}
  \left(\ln \frac{m^2_n}{Q^2} - \frac{3}{2} \right) ,
\end{equation}
where $m_n^2$ is the field-dependent mass-squared eigenvalue
of the $n$-th degree of freedom with spin $s_n$.
The renormalisation scale $Q$ should be chosen so that
it minimises $V^{(1)}$.
This means that the optimal scale is a function of $\phi$.
Taking this scale would be satisfying but technically involved.
For numerical computation, the scale shall be set equal to
the common diagonal component of the soft sfermion masses.
% which nearly coincides with $\sqrt{m_{\wt{t}_1} m_{\wt{t}_2}}$,
% the scale normally employed for electroweak symmetry breaking.
This is to be supplemented with error estimation.
Fortunately, the bounce consists of $\phi$
limited within a range not too far from the sfermion mass scale
\cite{Kusenko:1996jn}.
Therefore, $V^{(1)}$ does not diverge % grow indefinitely
even if the CCB minimum is taken to infinity.

% \begin{equation}
%   \begin{aligned}
%   V^{(1)} &= \sum_n (-1)^{2 s_n} (2 s_n + 1) \, h\bigl(m^2_n\bigr)
%   \\
%   h(x) &= \frac{x^2}{4}
%   \left(\ln \frac{x}{Q^2} - \frac{3}{2} \right)
%   \end{aligned}
% \end{equation}

One could estimate
the uncertainty due to ignoring loop corrections to the action
by evaluating the shift made by one-loop,
\begin{equation}
  \label{eq:diff 1L}
  \SEonel[\phibounce_\onel] - \SE[\phibounce] ,
\end{equation}
where $\SEonel$ is given by~(\ref{eq:SE}) with
$V$ replaced by the real part of the one-loop effective potential $\Vonel$
in~(\ref{eq:V1L}) \cite{Weinberg:1992ds},
and the functional argument of each action is its bounce.
% and $\phibounce_\onel$ is the bounce solution of $\SEonel$.
% and $\phibounce$ are
% the bounce solutions of $\SEonel$ and $\SE$, respectively.
% and $\Delta V \equiv \Vonel - V$.
One can approximate the difference~(\ref{eq:diff 1L}) by
\begin{equation}
  \label{eq:Delta S}
\begin{aligned}
  \DSE &\equiv \SEonel[\phibounce] - S[\phibounce]
  \\
  &= 2\pi^2 \int_0^\infty d\rho\rho^3\, \re[\Delta V (\phibounce(\rho))] ,
\end{aligned}
\end{equation}
where
\begin{equation}
  \label{eq:Delta V}
  \Delta V = \Vonel - V ,
\end{equation}
keeping only the leading order term in $\Delta V$.
% Truncating terms of the second order in $\Delta V$ and higher,
Note that $\Delta V$ is not simply equal to
$V^{(1)}/(16\pi^2) + \cosmoconst$
as may be suggested by~(\ref{eq:V1L}) since the parameters entering
$V$ and $\Vonel$ in~(\ref{eq:Delta V}) have different values.
In particular, those parameters pertinent to the electroweak symmetry breaking
must deviate.
Otherwise it would be impossible for both $V$ and $\Vonel$
to result in realistic Higgs vacuum expectation values.

\section{Procedure of analysis}
\label{sec:analysis}

% mass- and A- matrices
%% masses are diagonal
%% turn on one Aij at a time
% other parameters from softsusy
%% gauge couplings & yukawas at the scale of squark mass
%% mu b m2Hu m2Hd (\equiv mu mA tanb) set by hand
%%% avoid CCB directions not caused by FV Aij
%% M1 M2 M3 set by hand for the CW term

The method of numerical analysis is spelt out in detail.
The same procedure is repeated for each of the species $f = e,d,u$
and for each of the six flavour-off-diagonal trilinear couplings.

The families of the scalars $\wt{f}$
and $\wt{f}^c$ are defined in the super-CKM basis
where the corresponding Yukawa matrix $\lambda^f_{ij}$ is diagonal.
The CKM mixing is neglected so that
both $\lambda^u$ and $\lambda^d$ are diagonal at the same time.
This simplifies the computation as
the flavour-mixing Yukawa couplings % terms with couplings
% Yukawa couplings due to the CKM mixing
are dropped from the superpotential.
They are of $\mathcal{O}(V_{ts} \lambda_t) \sim \lambda^2$ or less,
and presumably are not very important in view of the fact that
a coupling of similar order
$\lambda_b \sim 0.1$ does not make a sizeable difference
as will be demonstrated later in this article.
%  seems a circular argument: should prove that CKM mixing does not
%  lower the potential energy
% Moreover, CKM mixing becomes irrelevant if
% only one-type of squarks and Higgs doublets have nonzero values
% as is the case for the set of parameters used here.

As the focus is on flavour-violation by the trilinear couplings,
the soft masses shall be assumed to be in a universal form,
\begin{equation}
  \label{eq:M2f}  
  (M^2_f)_{ij} = (M^2_{f^c})_{ij} = \msf^2\,\delta_{ij} .
\end{equation}
% \begin{equation}
%   \label{eq:M2f}
%   \begin{bmatrix}
%     \msf^2 & 0 & 0 \\ 0 & \msf^2 & 0 \\ 0 & 0 & \msf^2
%   \end{bmatrix}
% \end{equation}
Among the trilinear couplings, one $\Af{ij}$ with $i \neq j$
% flavour-off-diagonal component
is scanned with the step size of $0.1\TeV$ while all the other components
are kept zero.
The diagonal, of no interest, is set to zero.
One might well imagine a situation where more trilinears are turned on
in addition to $\Af{ij}$.
The additional couplings do not block an existing tunnelling path but
may open more \cite{Casas:1995pd}.
In this sense,
the bounds obtained in this work should be regarded as generous.

The phase of a single $\Ae{ij}$ can be rotated away by a
lepton-flavour number transformation.
Similarly, the phase of $\Ad{ij}$ or $\Au{ij}$ may be absorbed
into the squark fields by `quark-family number' transformation
in the limit of no CKM mixing that is taken here.

% Throughout the analysis,
The gauge and the Yukawa couplings
% The running parameters including the gauge couplings
are set at the renormalisation scale $Q = \msf$.
Therefore, $\Af{ij}$ presented later are also supposed to be at this scale.
%The following set of parameters shall be used unless stated otherwise:
The other parameters shall be set as follows unless stated otherwise:
\begin{equation}
  \label{eq:default}
  \begin{aligned}
    \tb = 10, \quad
    &\mu = 0.5\TeV, \quad
    \MH = 0.5\TeV, \\
    &\msq = \msl = 3\TeV .
  \end{aligned}
\end{equation}
The Higgs mass parameters, $m^2_{H_d}$, $m^2_{H_u}$, and $b$,
are determined from these by
the tree-level electroweak symmetry breaking condition.
More parameter space is also explored by varying each of the above.
A remark is in order regarding the sign of $\mu$.
The only non-negligible term affected by this in the scalar potential is
$2\,\re [\mu H_d \wt{Q}_3^* \lambda_t \wt{u}^{c*}_3]$.
Therefore, negating $\mu$ can be compensated for by flipping the sign of
either $\wt{Q}_3$ or $\wt{u}^c_3$.
Note that this can be done without touching the trilinear term in question.
In the case of a flavour-conserving trilinear, by contrast,
the signs of $\Au{33}$ and $\mu$ cannot be disentangled \cite{Casas:1995pd}.

For faster computation,
every scalar field is constrained to be real-valued.
This is not completely general in that
for a given set of real-valued $\{ \phi_\alpha \}$,
% $\min_{0 \le \theta_\alpha < 2\pi} V(e^{i \theta_\alpha} \phi_\alpha) \le
%  \min_{\theta_\alpha = 0,\pi } V(e^{i \theta_\alpha} \phi_\alpha)$
\begin{equation}
\min_{0 \le \theta_\alpha < 2\pi} V(e^{i \theta_\alpha} \phi_\alpha) \le
 \min_{\theta_\alpha = 0,\pi } V(e^{i \theta_\alpha} \phi_\alpha) ,
\end{equation}
even if all the parameters in $V$ are real.
One may worry that there can be a complex-valued path in the field space
that costs less potential energy.
Nevertheless, it can be checked that the equality holds
as long as $\{ \phi_\alpha \}$ stays sufficiently close
to one of the directions in~(\ref{eq:D-flat direction}),
and thus the potential is dominated by a few terms
including the activated trilinear term in particular.
It turns out that the numerical bounce does reveal this property.

Under the same condition, one can show that it costs less potential energy
for the squark colour directions to be aligned.
This justifies retaining only one colour component of each squark
assuming that the others are all zero.
% into the set of field variables.

As the problem is concerned with a flavour-off-diagonal coupling,
one should put at least two families of scalar fields into action
as variable degrees of freedom in addition to the Higgs doublets.
For the lepton sector, the following 10 variables are used:
$H_d, H_u, \wt{L}_i, \wt{e}^c_i$,
where the index $i$ runs over the two families involved.
Right-handed sneutrinos are not taken into account.
The quark sector needs two more variables:
$H_d, H_u, \wt{Q}_i, \wt{d}^c_i, \wt{u}^c_i$.

The bounce is computed numerically.
The Euclidean action~(\ref{eq:SE}) is approximated
by a function of the fields that are put on the discretised $\rho$ axis.
Each lattice point has an independent set of
the aforementioned 10 or 12 degrees of freedom.
One can find a stationary point constrained by
the boundary conditions~(\ref{eq:BCs}) using
a method derived from that in Ref~\cite{Konstandin:2006nd}.
The original series of steps had to be modified since it is not suitable
for a problem in which the true vacuum is very far from
the false vacuum in comparison to the thickness of the potential barrier.
This happens if the Yukawa couplings in~(\ref{eq:D-flat potential})
are small which is mostly the case except for $\lambda_t$.
The details of the revision are presented in a separate article
\cite{my method}.
With this method, one does not have to know
where the true minimum is, % or whether or not it exists
or even whether or not the potential is bounded from below.
The price to pay is that one needs to minimise a function
obeying constraints given in implicit forms.
This can be done for instance by using the Ipopt package
\cite{ipopt}.

At the first stage of the computation,
one needs to choose an initial field configuration.
% The above method requires a trajectory in the field space
% that starts from any point $\phib$ such that $V(\phib) = V(\phifv)$
% with $\phifv$ being the false vacuum
% and then goes over the barrier arriving at $\phifv$.
The above method needs a trajectory in the field space
that connects $\phifv$ to any point $\phieinit$
over the barrier such that $V(\phieinit) = V(\phifv)$.
To find a $\phieinit$, a point leaves the SML vacuum and hikes along a valley
in the direction of increasing $|\phi|$ until it goes down and
touches the initial altitude.

Finally, the SML vacuum lifetime is compared with the age of the universe
using~(\ref{eq:limit on S}), to accept or reject the given parameter set.

Apart from the scanning procedure described above,
error estimation is performed for a few selected cases
by evaluating the approximate one-loop effect~(\ref{eq:Delta S}).
The parameters entering the effective potential~(\ref{eq:V1L})
are generated by softsusy \cite{Allanach:2001kg}
with options set for one-loop Higgs masses and tadpoles.%
\footnote{There is a bug in softsusy version 3.1.7 that
discards even one-loop tadpoles when the option numRewsbLoops is set to 1.
This problem does not appear with the default setting.}
Unlike the tree-level potential, $\Vonel$ depends on
the gaugino masses as well.
They are picked up so that $M_{1,2,3} = \msf$
for order-of-magnitude estimation.

%\section{Results and discussions}
%\section{Results}
\section{Upper bounds and interpretations}
\label{sec:results}

% bounds for SE < 160, 250, 400, 630, 1000
% universality
%? analytic interpretation -> can't solve the DE
% radiative corrections
%% correction by Coleman-Weinberg term
%%% how big difference it makes
%% consequences of tree-level ewsb
%%%? compare with symmetric phase
%% scale dependence
%%% of bound on Aij
%%% of size of CW correction
% dependence on tanb, (sign of) mu, mA
% complex (sign of) Aij, multiple nonzero Aij, complex field values
% confirm: UFB direction irrelevant
% possible bound enhancement by presence of additional fields
% compare Nierste et al.
% empirical inequality a la Kusenko et al

\begin{figure}
  \centering
  \incgraph[height=\squarefiguresize]{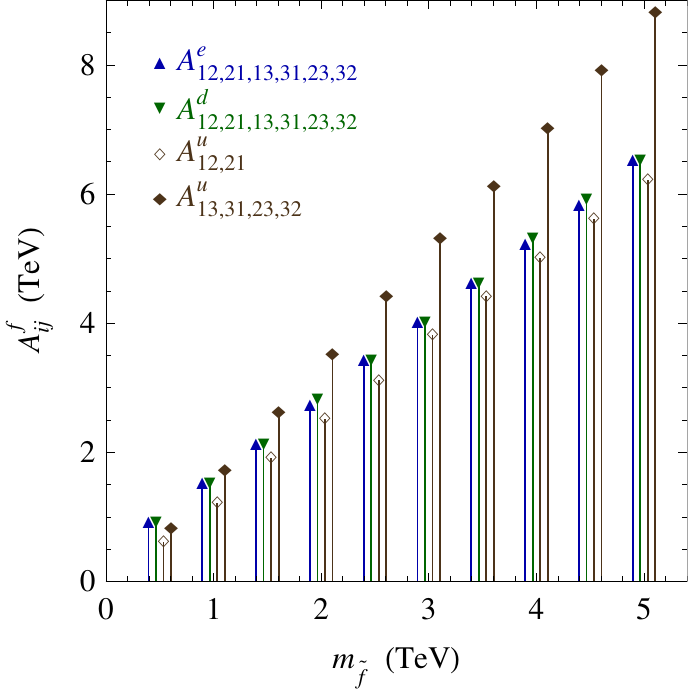}
  \caption{Upper bounds on flavour-violating $\Af{ij}$ as functions of $\msf$
  for $f = e,d,u$ and $i,j = 1,2,3$.
  They correspond to the following choice of parameters:
  $\tb = 10$,  $\mu = 0.5\TeV$,  $\MH = 0.5\TeV$.
  Each group of four vertical lines shares the same value of $\msf$
  given by the average of their horizontal coordinates.}
  \label{fig:mf dependence}
\end{figure}
Requiring that the bounce action obey the condition~(\ref{eq:limit on S}),
% be not less than 400,
one gets the upper bounds on the off-diagonal
trilinear couplings shown in Fig.~\ref{fig:mf dependence}.
It turns out that
the six components of $\Ae{ij}$ have almost the same bounds
(falling within 100 GeV from one another), and therefore
a single vertical line is used to display them all together.
The same is true of the six components of $\Ad{ij}$.
As for $\Au{ij}$, the upper bounds highly depend on whether
either of the family indices is 3 or not.
The bounds on $\Au{12}$ and $\Au{21}$ can be displayed together,
and likewise the other four couplings.

One can notice several features on the plot.
First of all, the upper bound on an off-diagonal $A$
grows as the corresponding sfermion mass is increased.
The dependence turns out to be almost linear although not exactly.
Second, the bound is symmetric under the interchange
of the two family indices.
Third, the bound on $\Ae{ij}$ is independent of $i$ and $j$
as mentioned above.
This family independence is discovered also in $\Ad{ij}$.
This property contrasts with the Yukawa dependence
of the stability bounds~(\ref{eq:stability bound on A}).
Fourth, the bound on $\Ae{ij}$ is nearly the same as that on $\Ad{ij}$.
Fifth, the bound on $\Au{12}$ is slightly lower
than that on $\Ade{ij}$.
Sixth, the bound on $\Au{13,23}$ is higher than that on $\Au{12}$
and also tend to be higher than the other bounds for
$\msf \gtrsim 1\TeV$.
Interpretations of these observations shall be given in what follows.

\begin{figure*}
  \centering
  \hspace{-7ex}
  \subfigure[\ $\Ae{23}$\hspace{-4ex}]
  {\incgraph[height=\squarefiguresize]{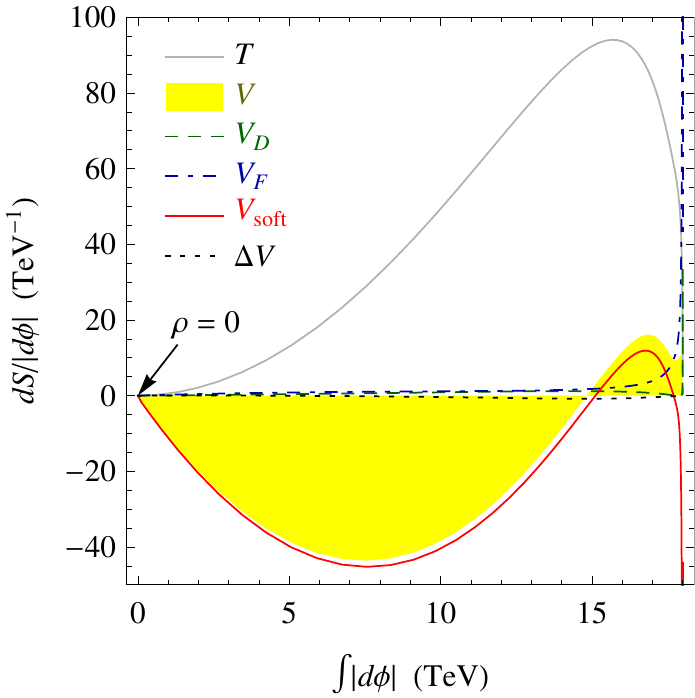}}\quad
% \subfigure[]{\incgraph[height=\squarefiguresize]{mixedSeffAd23}}\quad
  \subfigure[\ $\Au{12}$\hspace{-4ex}]
  {\incgraph[height=\squarefiguresize]{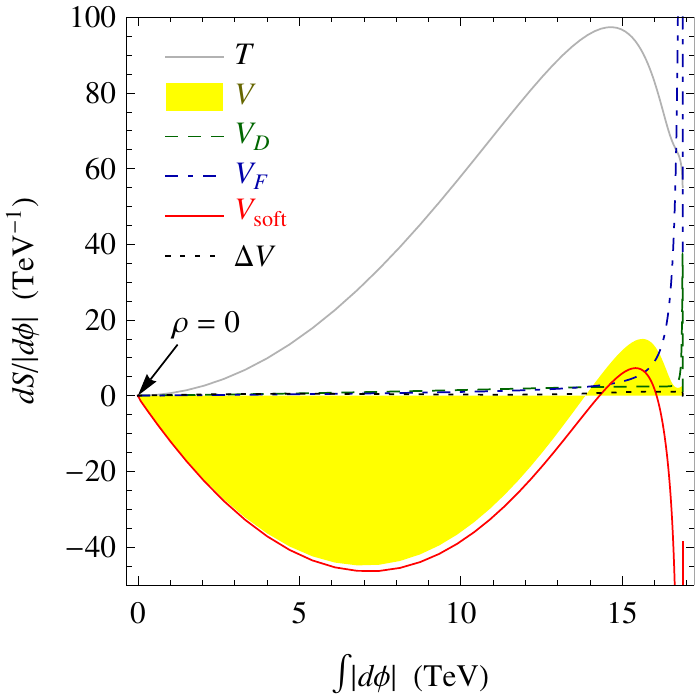}}\quad
  \subfigure[\ $\Au{23}$\hspace{-4ex}]
  {\incgraph[height=\squarefiguresize]{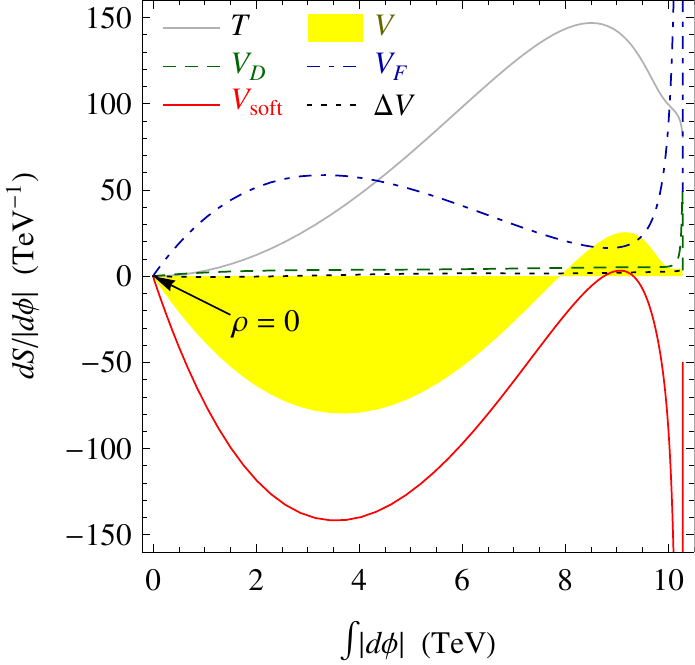}}
  \hspace{-5ex}
  \caption{Contributions of different potential components to the
    Euclidean bounce action.
    The nonzero trilinear coupling is (a) $\Ae{23} = 4\TeV$,
    (b) $\Au{12} = 3.8\TeV$, or (c) $\Au{23} = 5.3\TeV$.
    The (signed) light grey areas (yellow online)
    display the action coming from
    the total potential energy density $V$ that is the sum of
    $V_D$, $V_F$, and $V_\mathrm{soft}$, represented by
    the dashed (green online), the dot-dashed (blue online),
    and the solid grey (red online) curves, respectively.
    A black dotted curve is the estimated one-loop correction.
    The kinetic term is also displayed as a light grey curve.}
  \label{fig:decomposition}
\end{figure*}
For understanding the results,
it should be instructive to examine contributions from
different components of the scalar potential separately.
Three instances are shown in Figs.~\ref{fig:decomposition},
one with lepton flavour violation (LFV) and the other two with
up-type quark flavour change.
The figure for the down-quark sector is nearly identical to
Fig.~\ref{fig:decomposition}(a) and therefore has been omitted.
Each plot displays the differential action that arises from
each of $V_D$, $V_F$, $V_\mathrm{soft}$, and their sum denoted by $V$,
as one moves along the trajectory of the bounce profile.
The horizontal axis is the distance traversed in the field space
starting from the point of $\rho = 0$.
Every curve arrives on the right at the false vacuum.
Therefore, the area under (or over) each curve is the action
coming from the given potential component.
Also shown is the (positive-definite)
differential action from the kinetic term.
Obviously, the total Euclidean action is given by the sum of the area
under the light grey curve and the area of the light grey region
(yellow online).
% In Fig.~\ref{fig:decomposition}(a),
% an example with lepton-flavour violation is shown.
The parameters used in Figs.~\ref{fig:decomposition}
are as in~(\ref{eq:default}) besides
the trilinear couplings indicated in the caption.
% a leptonic trilinear coupling $\Ae{23} = 4\TeV$.
Each parameter set leads to the action $S \approx 400$,
that is, the $A$ coupling is around the metastability limit.

\begin{table}
  \centering
  \begin{tabular}{@{\hspace{3em}}l|r@{\extracolsep{1em}}r@{}r}
    \hline
    \multicolumn{1}{c|}{Panel in Figs.~\ref{fig:decomposition}}
                      &(a) $\Ae{23}$&(b) $\Au{12}$&(c) $\Au{23}$\\
    \hline
    $S_T$             &   781       &   763       &   749       \\

    $S_D$             &    12       &    22       &    41       \\
    $S_F$             &    39       &   116       &   529       \\
    $S_\mathrm{soft}$ & $-446$      & $-524$      & $-947$      \\
    $S_V$             & $-395$      & $-385$      & $-377$      \\

    $S$               &   386       &   378       &   372       \\
    \hline
    $\DSE$            &   $-6$      &    $8$      &   $11$      \\
    \hline
  \end{tabular}
  \caption{The Euclidean action $S$ decomposed into parts coming from
  the kinetic term, $V_D$, $V_F$, $V_\mathrm{soft}$, and $V$, respectively
  labelled $S_T$, $S_D$, $S_F$, $S_\mathrm{soft}$, and $S_V$,
  in each of the three cases shown in Figs.~\ref{fig:decomposition}.
  The last row shows the estimated one-loop effect.}
  \label{tab:decomposed S}
\end{table}
On each plot, the area under the $V_D$ curve is tiny
compared to that under $V$.
This means that $V_D$ does not play a very important role
in determining the action among the different components of
the total scalar potential $V$.
This point appears more concrete in Table~\ref{tab:decomposed S} which lists
the numerical value of each area.
Examining the field configuration, one can indeed find that
the trajectory of the solution almost follows one of the $D$-flat directions
in~(\ref{eq:D-flat direction}).
The same holds in every case considered in this work
in either the lepton or the quark sector.
This allows one to understand the qualitative features of
the metastability bound on a trilinear coupling by looking at
the potential~(\ref{eq:D-flat potential}).
One property of $V_\mathrm{L.E.}$ is that
it is symmetric under the interchange of $i$ and $j$ if
both $(M^2_\mathrm{av})^f_{ij}$ and $\Af{ij}$ are symmetric.
% $M^2_{f_i} = M^2_{f^c_j}$ and $\Af{ij} = \Af{ji}$.
This makes it understandable that the bound on $\Af{ij}$ is symmetric.
However, this apparent symmetry stems from the assumed form of
the sfermion mass matrix in~(\ref{eq:M2f}).
% what about non symmetric matrix?
Nevertheless, one could make a straightforward guess
at the behaviour for a more general mass matrix:
the bound should be a function of $(M^2_\mathrm{av})^f_{ij}$.
This property has been verified numerically.

The second smallest contribution comes from $V_F$.
It is totally negligible in a case for $\Ae{ij}$ or $\Ad{ij}$.
This explains why the upper bounds on $\Ae{ij}$ and $\Ad{ij}$ nearly coincide
in Fig.~\ref{fig:mf dependence}.
They do not depend on $i$ or $j$, and are even independent of
whether they are for charged sleptons or down-type squarks.
The only differences between the two species
are their gauge couplings
provided that their masses and trilinear couplings are equal
and that their Yukawa couplings are ignored.
Since both $S_F$ and $S_D$ are negligible, the Yukawa
and the gauge couplings do not make a noticeable difference between
the bounds on $\Ae{ij}$ and $\Ad{ij}$.
This should be clear from the potential in~(\ref{eq:D-flat potential})
with the Yukawa terms discarded.
As for $\Au{12}$, $S_F$ is not as small but
still minor compared to $S_\mathrm{soft}$.
Its bound in Fig.~\ref{fig:mf dependence} is quite close to
but a little stronger than those on $\Ade{ij}$, although
the charm Yukawa is small as well.
This split stems from the fact that
$\Au{12}$ is associated with the up-type Higgs unlike $\Ade{ij}$.
This point shall be elaborated on later.

A trilinear coupling including a stop
is allowed to be substantially larger than $\Au{12}$
even though they involve the same Higgs field,
as can be found in Fig.~\ref{fig:mf dependence}.
The reason is the large top Yukawa coupling.
It is evident in~(\ref{eq:D-flat potential}) that
Yukawas lift the scalar potential as quartic couplings,
and so they slow down the tunnelling process if they are large.
Therefore, large $\lambda_t$ implies weaker restrictions on $\Au{13,23}$.
The nontrivial role of $\lambda_t$ manifests itself
in both Fig.~\ref{fig:decomposition}(c) and the rightmost column of
Table~\ref{tab:decomposed S}
which show that $V_F$ is not negligible as for the other $A$ terms.

The last element to be explained in Table~\ref{tab:decomposed S}
is the estimate of one-loop effects, which was defined in~(\ref{eq:Delta S}).
The differential version is plotted in Figs.~\ref{fig:decomposition}.
The magnitude of $\DSE$ turns out to be roughly 3\%
of the tree-level $\SE$ in the case of $\Au{23}$ and is less significant
in the other two.
By full computation of the one-loop bounce $\phibounce_\onel$,
the validity of $\DSE$ as an approximation of the one-loop shift
in~(\ref{eq:diff 1L})
has also been checked.
The latter quantity is indeed of the same order as $\DSE$
for any case shown in the table.
In combination with the smallness of $\DSE$,
this should a posteriori support the reliability of an analysis employing
the tree-level effective potential.

\begin{figure*}
  \centering
  \incgraph[height=\rectangularfigureheight]{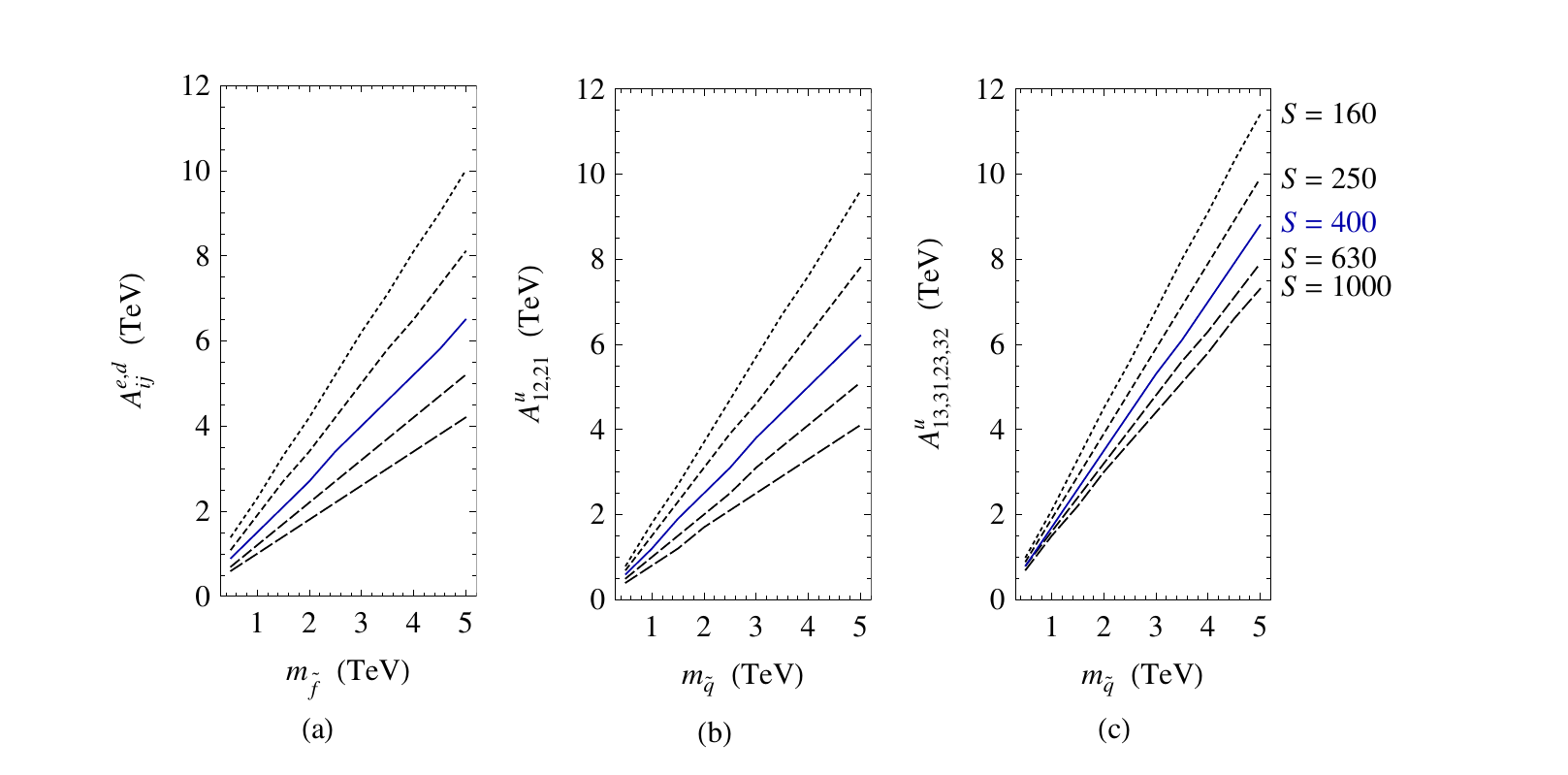}
  \caption{The plots show how the upper bounds change depending on
    the condition imposed on $\SE$.
    On each curve, the value of $\Af{ij}$ leads to the corresponding
    $\SE$ shown on the right,
    where $f = e,d,u$ and $i,j = 1,2,3$.
%  The longer a dash in a curve is, the larger the resulting $\SE$ is,
%  except the solid curves on which $\SE = 400$.
    The following values of parameters were used:
    $\tb = 10$,  $\mu = 0.5\TeV$, $\MH = 0.5\TeV$.}
  \label{fig:SE dependence}
\end{figure*}
Nevertheless, one should like to translate the variation
in $\SE$ into that in the maximum $\Af{ij}$.
In any case, there are non-vanishing uncertainties coming from
loop corrections as well as from the ignorance of the prefactor.
Also, one could choose to impose a lower limit on $\SE$
slightly different from that in~(\ref{eq:limit on S}).
All these variances could be encoded as a constant added to $\SE$.
Then, its influence on a trilinear coupling can be found
in Figs.~\ref{fig:SE dependence}.
If one allows for a change in $\SE$ by a factor between 0.6 and 1.6
for instance, the bound on $\Af{ij}$ is altered by % $^{+29}_{-22}\%$,
$^{+30}_{-20}\%$,
and each metastability bound presented later is weakened or strengthened
by the same fraction.

As is clear in Fig.~\ref{fig:mf dependence},
the tunnelling process gets slower
as the squark and the slepton masses increase
since they lift the potential~(\ref{eq:D-flat potential}) quadratically.
By the same token, one may envisage
a similar effect from
% that the tunnelling constraints should loosen for
higher masses of Higgs fields,
as either of them also acquires a vacuum expectation value along
the CCB direction~(\ref{eq:D-flat direction}).
Concerning the Higgs bosons,
the parameters directly linked to the tunnelling
are not the soft masses appearing in~(\ref{eq:D-flat potential})
(which can be tachyonic)
but the physical mass eigenvalues which describe the positive
curvature of the potential around the SML vacuum.
% as the squark and the slepton soft masses do.
In the MSSM, one can specify the tree-level Higgs potential
by choosing $\tb$, $\MH$, and $\mu$.
Once the former two parameters fix the complete Higgs mass spectrum,
varying $\mu$ makes no difference therein.
The dependence on each of these three parameters shall be discussed below.

\begin{figure*}
  \centering
%  \subfigure[\ $\Ax{e,d}_{ij}$\hspace{-4ex}]
  \subfigure[\hspace{-4ex}]
  {\incgraph[height=\squarefiguresize]{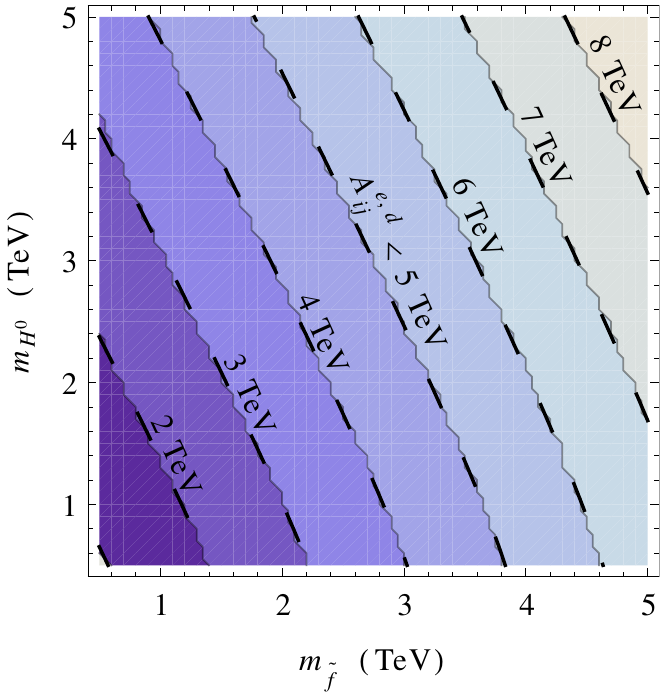}}\quad
%  \subfigure[\ $\Au{12,21}$\hspace{-4ex}]
  \subfigure[\hspace{-4ex}]
  {\incgraph[height=\squarefiguresize]{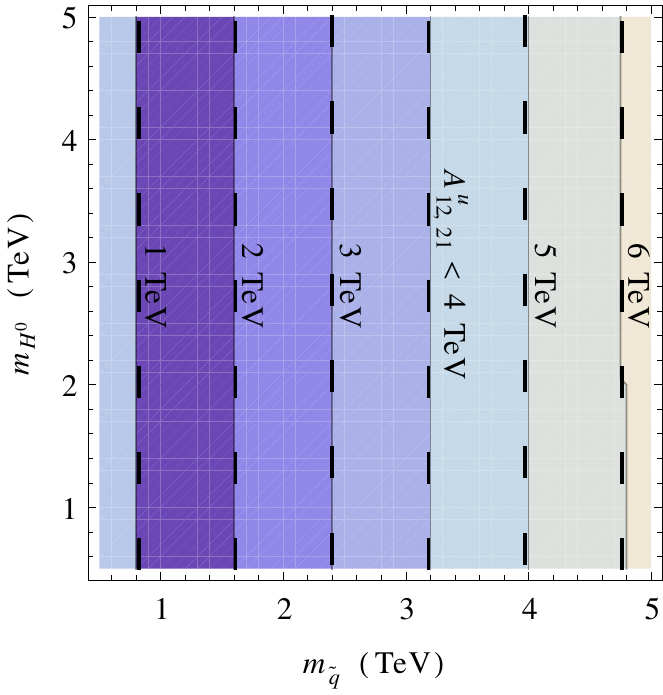}}\quad
%  \subfigure[\ $\Au{13,31,23,32}$\hspace{-4ex}]
  \subfigure[\hspace{-4ex}]
  {\incgraph[height=\squarefiguresize]{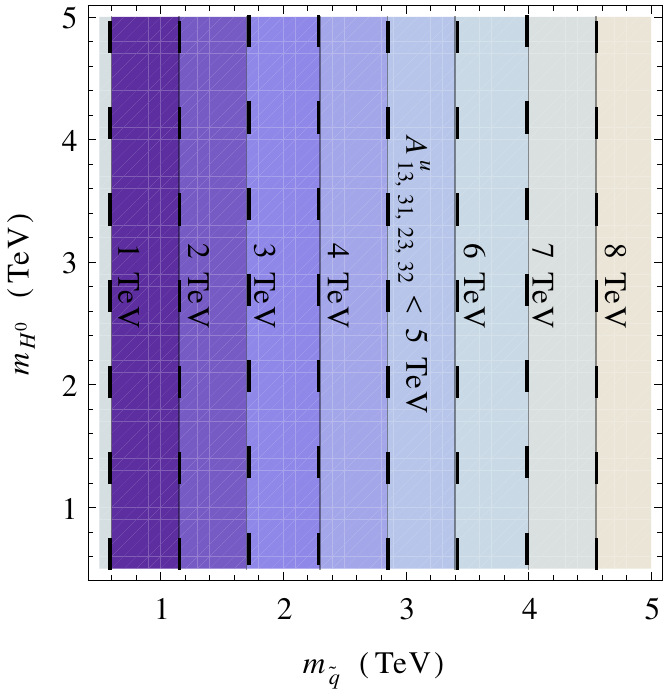}}
  \caption{Contours of upper bounds on flavour-violating $\Af{ij}$
    on the $(\msf,\MH)$ plane
    for $f = e,d,u$ and $i,j = 1,2,3$.
    The grey contours [which look wiggly in Fig.~(a)]
    are based on the data from grid scanning
    and the dashed black curves are the least-squares fits.
    They correspond to the following choice of parameters:
    $\tb = 10$,  $\mu = 0.5\TeV$.}
  \label{fig:contours}
\end{figure*}
First, $\MH$ and the sfermion masses are varied
while the other parameters are fixed according to~(\ref{eq:default}).
The results are set out in Figs.~\ref{fig:contours} as grey contours.
As expected, enlarging $\MH$ does provide more room for
the trilinear couplings.
However, the effect is significant only on those
involving the down-type Higgs, displayed in Fig.~\ref{fig:contours}(a).
Contours of maximum $\Au{ij}$ in Figs.~\ref{fig:contours}(b) and (c),
do not exhibit a noticeable dependence on $\MH$.
One can understand this by considering Higgs mixing.
The real part of each Higgs field can be expressed as
(see e.g.\ Ref.~\cite{Martin:1997ns})
\begin{subequations}
  \begin{align}
  \re H_u^0 \approx v_u + ( \sin\!\beta\, h^0 - \cos\!\beta\, H^0 )/\sqrt{2} ,
  \\
  \re H_d^0 \approx v_d + ( \sin\!\beta\, H^0 + \cos\!\beta\, h^0 )/\sqrt{2} ,
  \label{eq:Hd0}
  \end{align}
\end{subequations}
in terms of the mass eigenstates.
The Higgs mixing angle $\alpha$ has been approximated by $\beta - \pi/2$
since $\MH \gg m_Z$ in most of the parameter space considered in this work.
For $\tb = 10$, $H_u^0$ and $H_d^0$ are mostly $h^0$ and $H^0$, respectively.
Since the mass of the lighter $CP$-even neutral Higgs boson is always
restricted around $m_Z$, the potential curvature along the
$H_u^0$ direction stays essentially constant
in comparison to the large variation in the squark mass.
This property should remain true at least qualitatively for any
$\tb$ bigger than a few.

A metastability bound should be easier to use if one has
its algebraic expression.
Within the parameter region shown in the figures,
the bounds can be approximated by the empirical inequalities,
\begin{subequations}
  \begin{align}
    \begin{split}
    \Ade{ij} <&\
%    0.613028 \MH + 1.24849 \msf - 0.107941 \sqrt{\MH \msf} + 0.0387731
%    0.61\,\MH + 1.25\,\msf \\
%    &- 0.11 \sqrt{\MH \msf} + 39 \GeV ,
%    0.0369427\[VeryThinSpace]+ 0.592196 mH - 0.452948 mH^0.0654126 ml^0.934587 + 1.61458 ml
    1.61 \,\msf + 0.59 \,\MH \\
    &- 0.45 \,\msf^{0.93}\MH^{0.07} + 37\GeV ,
    \end{split}
%    0.61 \MH &+ 1.25 \msf - 0.11 (\MH \msf)^{\frac{1}{2}} + 39 \GeV
     \\
%    \Au{12,21} &< 1.26948 \msq - 0.0439475
     \Au{12,21} <&\ 1.27\,\msq - 44 \GeV ,
     \\
%    \Au{13,31,23,32} &< 1.76428 \msq - 0.0324418
     \Au{13,31,23,32} <&\ 1.76\,\msq - 32 \GeV .
  \end{align}
\end{subequations}
These are the least-squares fits to the numerical data
using the form, $a\,\msf + b\,\MH + c\,\msf^p \,\MH^{1 - p} + d$.
% The others were fit with $c = 0$.
They are plotted as dashed black curves in Figs.~\ref{fig:contours}
which reveal that the fits are indeed faithful approximations of the data.
These inequalities should remain nearly the same
even for different values of $\tb$ and $\mu$
since the bounds are insensitive to them as will be shown below.
It should also be remembered that
$\msf^2\,$ should be replaced by $(M^2_\mathrm{av})^f_{ij}$
if the diagonal soft masses are highly nondegenerate.

\begin{figure*}
  \centering
\incgraph[height=\rectangularfigureheight]{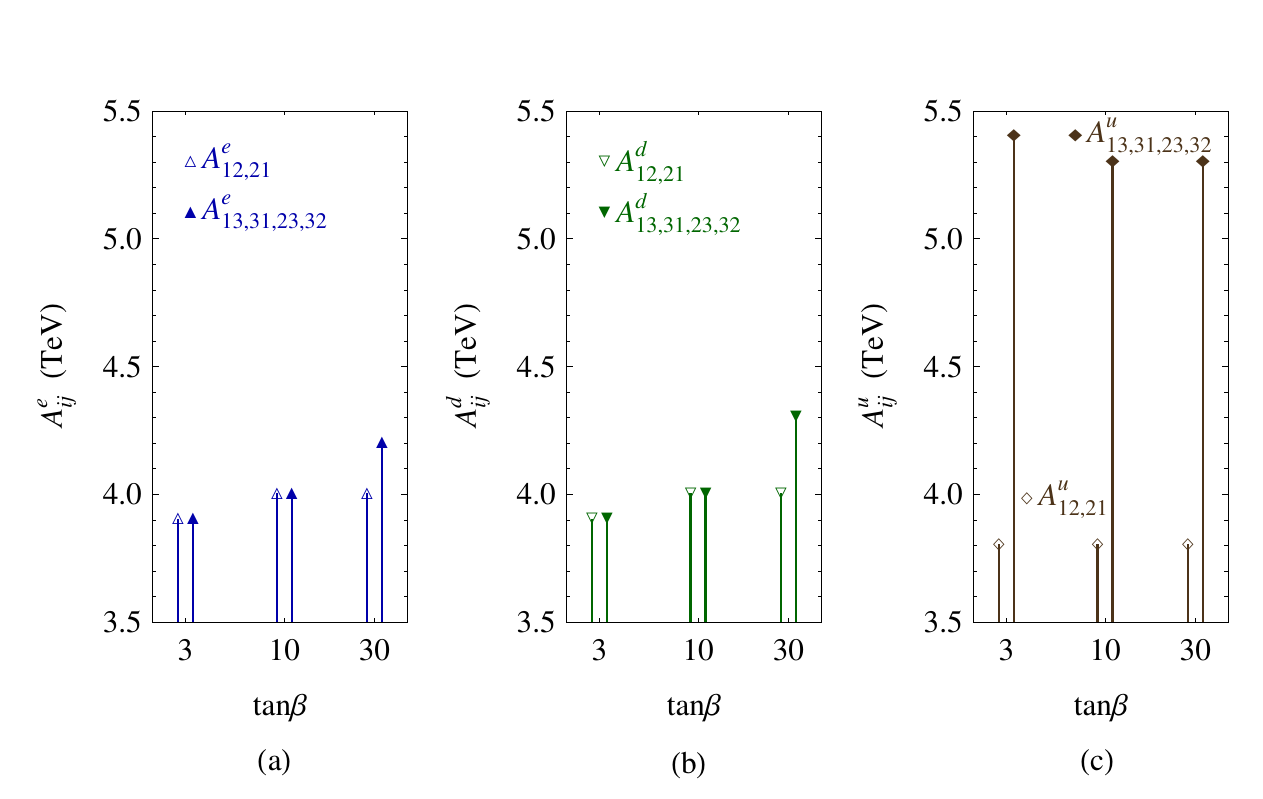}
  \caption{Upper bounds on flavour-violating $\Af{ij}$ as functions of $\tb$
    for $f = e,d,u$ and $i,j = 1,2,3$.
    They correspond to the following choice of parameters:
    $\mu = 0.5\TeV$, $\MH = 0.5\TeV$, $\msf = 3\TeV$.
    Each pair of vertical lines shares the same $\tb$.}
  \label{fig:tanb dependence}
\end{figure*}
Next, the dependence on $\tb$ is illustrated
in Figs.~\ref{fig:tanb dependence}.
There are two origins of change as $\tb$ varies.
One is the Higgs mixing mentioned above.
This works in the same way on all the species
that couple to each type of Higgs.
As $\tb$ increases, $H_d^0$ in~(\ref{eq:Hd0})
acquires more fraction of the heavier Higgs
which impedes the tunnelling caused by $\Ade{ij}$.
This effect appears as the slight rise of the $\Ade{12}$ points for
higher $\tb$ in Figs.~\ref{fig:tanb dependence}(a) and (b).
For $\Au{ij}$, the change is in the opposite way but too small to be visible
on the points for $\Au{12}$ in Fig.~\ref{fig:tanb dependence}(c).
Remember that increasing $\MH$ renders the tunnelling less probable.
Therefore, the influence of Higgs mixing should be enhanced by higher $\MH$.
The other origin is the variation of Yukawa couplings,
which dominates the $\tb$ dependence of
the trilinear couplings involving the third family.
A higher value of $\tb$ implies
larger down-type quark and charged lepton Yukawas.
This gives rise to slower tunnelling as explained before.
% Note also that
This splits to some extent
the upper bounds that were close together for lower $\tb$.
The up-type Yukawas do not change as much as the down-type ones,
and this leads to the milder dependence of $\Au{13,23}$
in Fig.~\ref{fig:tanb dependence}(c).
After all, the alteration is at most 10\%, found in $\Ad{23}$,
as $\tb$ is increased from 3 to 30, which does not seem very significant.

% \begin{figure*}
%   \centering
% \incgraph[height=\rectangularfigureheight]{dep_mu}
%   \caption{Upper bounds on flavour-violating $\Af{ij}$ as functions of $\MH$
%     for $f = e,d,u$ and $i,j = 1,2,3$.
%     They correspond to the following choice of parameters:
%     $\tb = 10$, $\msf = 3\TeV$.
%   As for $\mu$, two values have been tried: $0.5\TeV$ and $3.0\TeV$.
%   Each pair of vertical lines shares the same $\MH$
%   given by the average of their horizontal coordinates.}
%   \label{fig:mu dependence}
% \end{figure*}

Finally, one should discuss the dependency on $\mu$.
As already stated, the Higgs mass eigenvalues and mixing angle
do not rely on $\mu$ after $\tb$ and $\MH$ have been fixed.
Therefore, the metastability bounds should remain the same
while $\mu$ is altered.
This has been checked by trying two values of $\mu$
of different orders of magnitude: $0.5\TeV$ and $3.0\TeV$.
For each $\mu$, $\MH$ has been scanned from $0.5\TeV$ to $5.0\TeV$,
with the other parameters set as~(\ref{eq:default}).
At these parameter points, the two choices of $\mu$
do lead to identical upper bounds.

%\section{Implications for existing and forthcoming searches}
\section{Implications for flavour physics}
\label{sec:comparison}

Having obtained the limits on flavour-violating trilinear couplings,
one should consider what consequences they have for
physical processes.
% In what follows,
% the metastability bounds shall be contrasted with other existing constraints
% primarily from FCNC processes, and prospects for
% future new physics searches therein be discussed.
In what follows,
the metastability bounds shall be contrasted with other existing constraints.
They are based primarily on new physics searches within FCNC processes,
of which prospects at running and planned experiments are also discussed.
Other indirect bounds are compared together
such as those coming from the vacuum stability, the $\rho$ parameter,
and naturalness of the CKM matrix.
In this section, the degree of flavour-violation is expressed
as a $\delta$ parameter, defined in~(\ref{eq:delta}),
that is normally used for flavour physics.

\begin{table}
  \centering
  \begin{tabular}{c|cc}
    \hline
    Mode   &
    \multicolumn{1}{c}{Current bound} &
    \multicolumn{1}{c}{Future bound} \\
    $\meg$ &
    $1.2\cdot 10^{-11}$ \cite{Brooks:1999pu}&
    $1  \cdot 10^{-13}$ \cite{MEG} \\
    $\teg$ &
    $3.3\cdot 10^{-8\phantom{1}}$ \cite{tegtmg}&
    $2  \cdot 10^{-9\phantom{1}}$ \cite{Bona:2007qt}\phantom{8} \\
    $\tmg$ &
    $4.4\cdot 10^{-8\phantom{1}}$ \cite{tegtmg}&
    $2  \cdot 10^{-9\phantom{1}}$ \cite{Bona:2007qt}\phantom{8} \\
    \hline
  \end{tabular}
  \caption{Present and future experimental sensitivities to
    the branching ratios
    of the lepton flavour violating decay modes that probe
    $\dee{ij}{LR}.$}
  \label{tab:lfv experiments}
\end{table}
The strongest experimental constraints on the $LR$ mass insertions
in the slepton sector come from radiative LFV decays.
The restrictions on their branching ratios are collected in
Table~\ref{tab:lfv experiments}.
The current upper limits are all at 90\% confidence level (CL).
The quoted future sensitivity to $\meg$ is
anticipated at 90\% CL from the MEG experiment \cite{MEG}.
According to Ref.~\cite{Bona:2007qt},
a SFF is expected to provide
the displayed 90\% CL bounds on $\teg$ and $\tmg$, whereas
Ref.~\cite{Aushev:2010bq} projects the $\tmg$ bound of $3\cdot 10^{-9}$.

Each bound is translated into that on the corresponding mass insertion
and presented below.
For this, one should choose the other parameters on which
the LFV decay rate depends.
In the case of $\dee{ij}{LR}$ or $\dee{ij}{RL}$ with $j > i$,
fixing the bino mass $M_1$ is enough
to determine $\BR(l_j \rightarrow l_i \gamma)$
in the mass insertion approximation \cite{Ciuchini:2007ha}.
As a demonstration, it shall be assumed that $M_1 = \msl$.

\begin{table}
  \centering
  \begin{tabular}{c|lc}
    \hline
    MI &
    \multicolumn{1}{c}{Current bound} &
    \multicolumn{1}{c}{Future bound} \\
    $\ded{12}{LR,RL}$ &
    $1.4\cdot 10^{-3}$ \cite{Buchalla:2008jp}
    \\
    $\ded{13}{LR,RL}$ &
    $2.9\cdot 10^{-2}$ \cite{Ko:2002ee,Buchalla:2008jp}&
    $2  \cdot 10^{-3}$ \cite{Bona:2007qt} \\
    $\ded{23}{LR,RL}$ &
    $1.4\cdot 10^{-2}$ \cite{Buchalla:2008jp} &
    $5  \cdot 10^{-3}$ \cite{Bona:2007qt} \\
    $\deu{12}{LR,RL}$ &
    $1.6\cdot 10^{-2}$ \cite{Buchalla:2008jp} \\
    \hline
  \end{tabular}
  \caption{Experimental sensitivities to flavour-violating mass insertions
    in the squark sector for $\msq = \mgl = 1\TeV$.
    Expectations of future searches are quoted if available.}
  \label{tab:dd constraints}
\end{table}
In the squark sector, there are multiple observables
that restrict a single mass insertion.
These constraints are combined to give the 95\% CL upper bound
on the modulus of each mass insertion in Table~\ref{tab:dd constraints}.
Also presented are the moduli that in the future
can be reconstructed at the level of $3\sigma$.
They are based on the estimates in Ref.~\cite{Bona:2007qt}.
An estimate in this reference is in the form of a range and the values
in Table~\ref{tab:dd constraints} are those at the lower ends,
i.e.\ the most optimistic ones.
The limit depends on the squark and the gluino masses as
it arises from gluino-squark loops.
The numerical values shown in the table are for $\msq = \mgl = 1\TeV$,
and they are proportional to the squark mass
as long as $\msq/\mgl$ is fixed.
This ratio shall be maintained in the following plots.

The FCNC constraints presented below should be regarded as rough estimates.
They rely on parameters that do not affect a metastability bound.
Already at one-loop level, they depend on gaugino masses
which have been chosen somewhat arbitrarily as stated above.
Moreover, chirally-enhanced higher order corrections can
significantly strengthen or weaken a flavour constraint
\cite{Crivellin:2009ar}.
These corrections shall be ignored that depend on yet more parameters.

% differences btw metas and CCB bounds
%% max(i,j) dependence
%% tan b dependence

\begin{figure*}
  \centering
  \subfigure[]{\incgraph[height=\squarefiguresize]{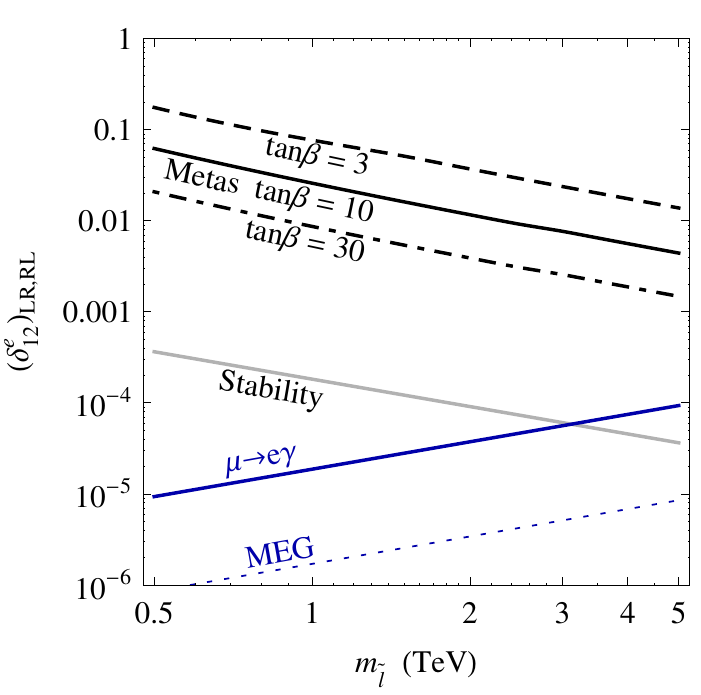}}\qquad
  \subfigure[]{\incgraph[height=\squarefiguresize]{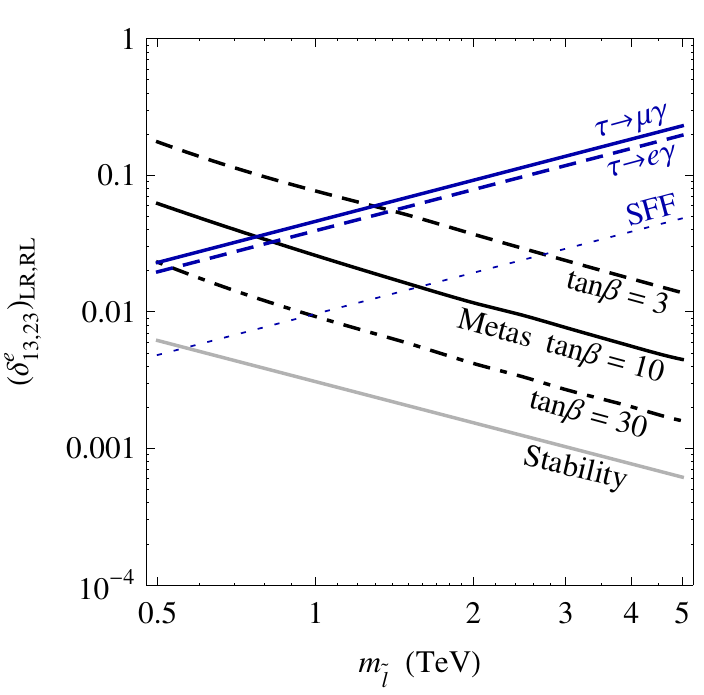}}
  \caption{Comparison of constraints on $\dee{ij}{LR}$.
  The thick black lines with negative slopes are bounds from the requirement
  that $\SE > 400$, for three values of $\tb$
  represented by different dashing styles.
  The thick grey lines are the stability bounds.
  The thick dark grey lines (blue online)
  with positive slopes are the current bounds
  from flavour-changing processes.
  The thin dotted lines (blue online) are anticipated experimental bounds.}
  \label{fig:de comparison}
\end{figure*}
% The first species to be considered is the charged leptons.
Given the aforementioned experimental information,
one could start the comparative analysis by examining the charged leptons.
In Figs.~\ref{fig:de comparison},
the vacuum bounds are displayed along with
the present and the projected sensitivities of FCNC searches.
The vacuum stability border is given by~(\ref{eq:stability bound on d}),
above which the corresponding trilinear term gives rise to tunnelling.
If one abandons the requirement that the SML vacuum be absolutely stable,
then the limit is relaxed up to the line
above which the false vacuum decays too fast in comparison to
the age of the universe.
As the sfermion masses increase,
both the stability and the metastability bounds become stronger.
However, the latter depends on $\tb$ unlike the former.
As Fig.~\ref{fig:tanb dependence}(a) shows, maximum $\Ae{ij}$
allowed by the vacuum lifetime is insensitive to $\tb$.
Therefore, (\ref{eq:delta}) implies a tighter restriction
on $\dee{ij}{LR}$ for higher $\tb$.
In contrast, the FCNC constraints loosen as the sfermions grow heavier.

More specifically,
Fig.~\ref{fig:de comparison}(a) is concerned with
the $\mu \rightarrow e$ transition.
Within the displayed range,
the tunnelling process does not play a significant role
in comparison to the $\meg$ decay.
The MEGA experiment has ruled out more region than metastability
so long as $\msl \lesssim 20\TeV$ for $\tb = 30$.
There is even no probability of tunnelling for up to
$\msl \approx 3\TeV$, beyond which the SML vacuum lifetime is
still permitted to be either finite or infinite.
% Therefore,
Allowing for an arbitrary mass insertion
consistent with each vacuum bound,
one can estimate what slepton mass range may be accessible through $\meg$.
% hope to discover $\meg$.  % at the MEG experiment.
The stability condition sets the higher end around $10\TeV$.
% The stability border crosses the MEG curve much earlier around $10\TeV$.
If one opens up the possibility of a long-lived false vacuum,
the territory extends up to $\msl \sim 60\TeV$ where
the metastability bound for $\tb = 30$ crosses the MEG sensitivity.
As is clear in the plot, this crossing point depends on $\tb$,
and is about $100, 200\TeV$ for $\tb = 10,3$, respectively.
Above this mass scale, the mass insertion is suppressed too much
to be observed.
The metastability bound has been numerically computed
only up to $\msl = 5\TeV$,
and the preceding estimates are based on linear extrapolation.

In a similar way, one can interpret Fig.~\ref{fig:de comparison}(b)
which shows the 1--3 and the 2--3 mixings at the same time.
Obviously, they are associated with $\teg$ and $\tmg$, respectively.
In contrast to $\meg$, their current experimental bounds
are less powerful and leave ample room for vacuum metastability
even for slepton masses as low as $500\GeV$.
If one were to insist on permanent stability,
it could be an explanation for why LFV in $\tau$ decays has not been
discovered yet.
This requirement would rule out even the possibility of observing
$\teg$ or $\tmg$ at a SFF, provided that
they arise solely from the trilinear terms and
that $\msl \gtrsim 600\GeV$.
Adopting the vacuum longevity constraint instead
makes the LFV search more promising.
A SFF might be able to find the virtual effect of
a $LR$ insertion involving the third family
% $\dee{13}{LR,RL}$ or $\dee{23}{LR,RL}$
if sleptons are lighter than $1.0,1.6,2.7\TeV$
for $\tb = 30,10,3$, respectively.
If one wishes to use the projected $\tmg$ bound of $3\cdot 10^{-9}$
from Ref.~\cite{Aushev:2010bq},
one may shift up the corresponding line in the plot
by the factor of $\sqrt{3/2}$.

\begin{figure*}
  \centering
\subfigure[]{\incgraph[height=\squarefiguresize]{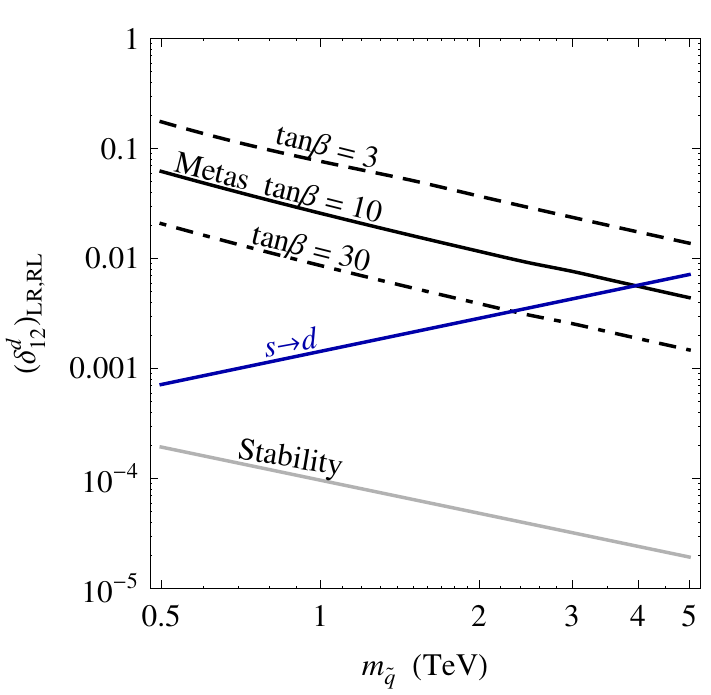}}\qquad
\subfigure[]{\incgraph[height=\squarefiguresize]{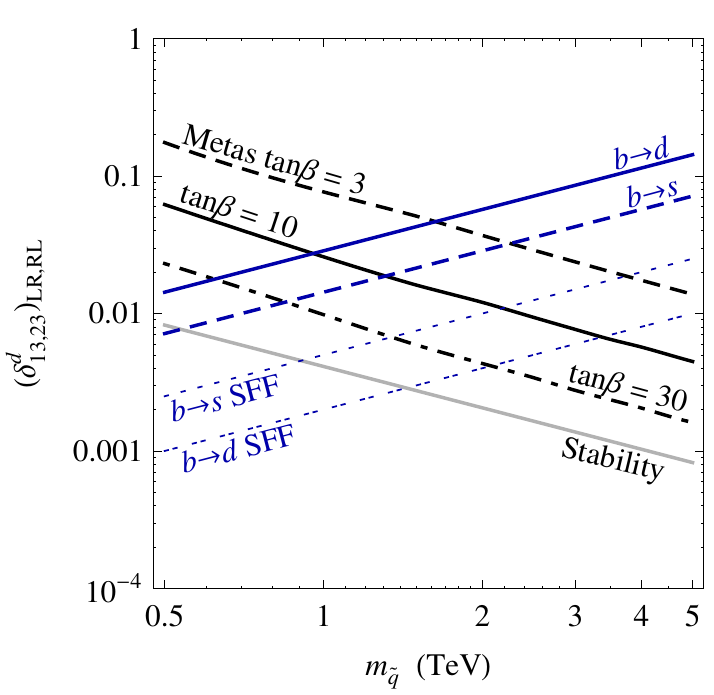}}
  \caption{Comparison of constraints on $\ded{ij}{LR}$.
  The thick black lines with negative slopes are bounds from the requirement
  that $\SE > 400$, for three values of $\tb$
  represented by different dashing styles.
  The thick grey lines are the stability bounds.
  The thick dark grey lines (blue online)
  with positive slopes are the current bounds
  from flavour-changing processes.
  The thin dotted lines (blue online) are anticipated experimental bounds.}
  \label{fig:dd comparison}
\end{figure*}
Next, the down-type quark sector is investigated.
The structure of each of Figs.~\ref{fig:dd comparison}
is the same as a leptonic one.
The left panel is devoted to the $s \rightarrow d$ transition.
The kaon physics constraint is weaker than the stability condition
but stronger than metastability up to the mass scale around a few TeV\@.
Again, the crossing position varies according to $\tb$ which can be
found in Fig.~\ref{fig:dd comparison}(a).
If squarks are heavier than this mass scale,
$\ded{12}{LR,RL}$ are bounded to be smaller than
the sensitivity of present measurements.
However, one should keep in mind that kaon dynamics is plagued by
large theoretical uncertainties and so is the experimental limit.

The mass insertions relevant to the bottom quark are
shown in Fig.~\ref{fig:dd comparison}(b).
The features of the 1--3 and the 2--3 sectors are similar,
with differences arising from the FCNC constraints.
The current $B$ physics data does not give much information with respect to
whether or not a vacuum decay can be
triggered by the related trilinear term.
As long as one stays within the stable region,
the scope of a SFF stops around $1.4\TeV$ via the 1--3 mixing
and a little lower via 2--3.
If one goes beyond the stability limit,
% up to the limit set by the age of the universe,
the reach of new physics search in $b \rightarrow d$ transitions
is raised up to $2.1, 3.4, 5.9 \TeV$ for $\tb = 30,10,3$, respectively.
The respective $b \rightarrow s$ counterparts are $1.3, 2.2, 3.7\TeV$.
As before, these limits are set by the age of the universe.
Remember that there are hadronic uncertainties
in heavy flavour physics as well
albeit less than in the kaon sector.
Also, the future bounds in Table~\ref{tab:dd constraints}
have rather big errors.
In particular, new physics is much easier to unveil
if there is an extra complex phase, since
$CP$ violation is less affected by hadronic uncertainties in general.
These points should be taken into account
when one interprets the above numerical results.

Before moving on to the up sector, it should be appropriate to
recall the Higgs mass dependence.
The preceding graphs correspond to the choice $\MH = 0.5\TeV$.
As demonstrated in Fig.~\ref{fig:contours}(a),
higher $\MH$ makes bigger $(\deltax{e,d}_{ij})_{LR}$ acceptable.
This can extend the boundary of reachable mass upwards.
% Multiplying an $LR$ insertion by factor 1.5 for instance
% could extend the boundary of reachable mass upwards by 20\%.

\begin{figure*}
  \centering
\subfigure[]{\incgraph[height=\squarefiguresize]{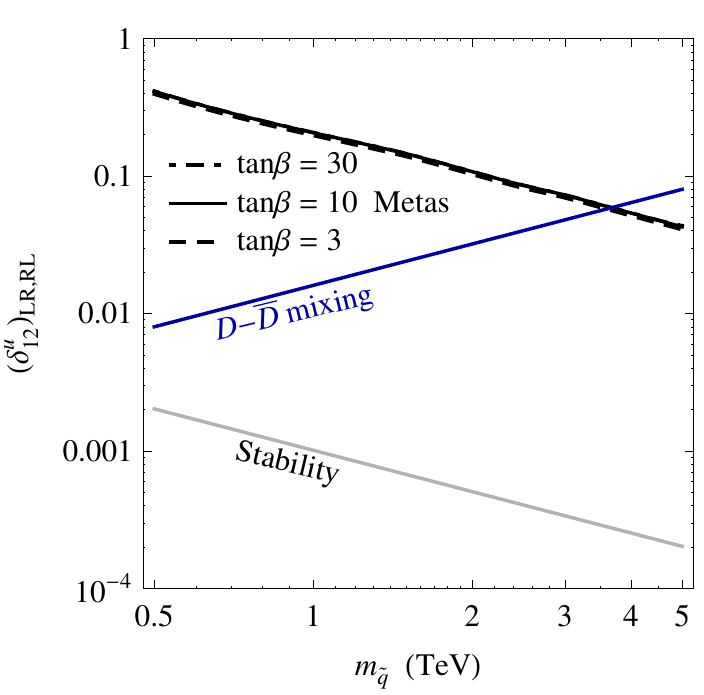}}\qquad
\subfigure[]{\incgraph[height=\squarefiguresize]{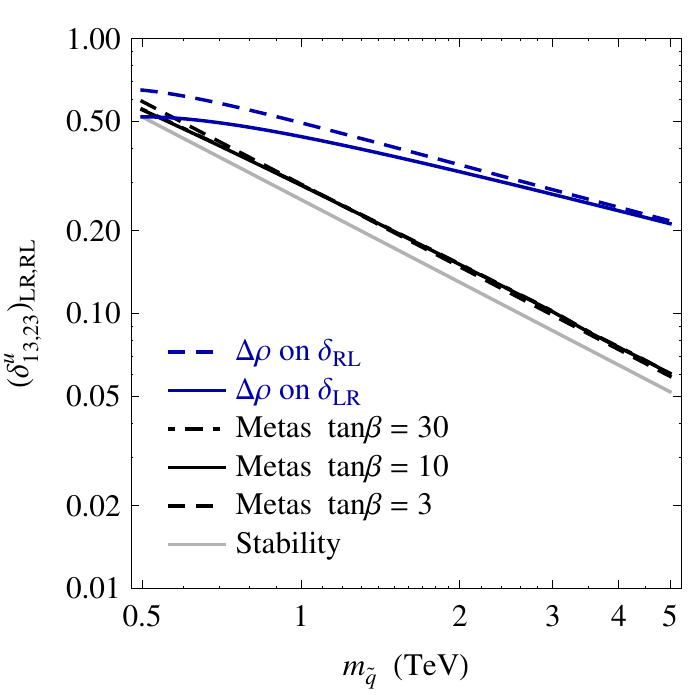}}
  \caption{Comparison of constraints on $\deu{ij}{LR}$.
  The thick black lines are bounds from the requirement
  that $\SE > 400$, for three values of $\tb$
  represented by different dashing styles.
  The thick grey lines are the stability bounds.
  The thick dark grey line (blue online) in panel (a)
  is the current bound from $D^0$--$\overline{D^0}$ mixing.
  In panel (b), the thick dark grey lines (blue online) are
  the bounds from the $\rho$ parameter.}
  \label{fig:du comparison}
\end{figure*}
Finally, Figs.~\ref{fig:du comparison} are allocated to
flavour violation of the up-type squarks.
One finds an outstanding distinction between
the metastability bounds shown here and in the preceding plots, i.e.\
the way they depend on $\tb$.
This is because the up-type squarks couple to $H_u$
unlike the fields considered above.
According to (\ref{eq:delta}),
the bound becomes looser as $\tb$ increases
although the dependence is much milder.

Regarding the 1--2 sector,
there is a constraint from a flavour-changing process.
In Fig.~\ref{fig:du comparison}(a),
the $D^0$--$\overline{D^0}$ mixing bound is plotted.
It does not forbid the SML vacuum having a finite lifetime,
but does restrict the tunnelling rate below the acceptable level
for $\msq \lesssim 4\TeV$.
Beyond this mass scale, the metastability bound is stricter
than the current data.

There is not much experimental information on the $LR$ insertions
related to the top squark.
In particular, no FCNC constraint is available yet.
Instead, Fig.~\ref{fig:du comparison}(b) employs
another indirect limit coming from electroweak precision observables.
It is obtained by requiring that
the squark contribution to the $\rho$ parameter \cite{Heinemeyer:2004by}
do not exceed $5.5\cdot 10^{-4}$ \cite{Cao:2007dk}.
It is based on the assumption that
$\ded{ij}{LR}$ is not correlated with $\deu{ij}{LR}$ so that
a large value of the latter breaks the custodial symmetry
between the up- and down-type squarks significantly.
As squarks grow heavier,
this limit does not decouple either but tightens.
However, the pace is slower than those of the vacuum constraints.
As a result, the vacuum longevity bound is more stringent
in most of the parameter space.
% claim this bound is strongest?
Another point to notice is that the stability and
the metastability curves are very close to each other
compared to the foregoing cases that do not involve the top quark.
This is mainly because the large top Yukawa coupling
renders the inequality~(\ref{eq:stability bound on A})
less restrictive.
In the plot is shown even a small region which is ruled out by
the tunnelling constraint for some $\tb$ even though it satisfies
the stability requirement.
Of course, a classically stable vacuum cannot decay.
This apparent nonsense stems from the fact that
the condition~(\ref{eq:stability bound on A}), used for drawing the curve,
is not optimal \cite{Casas:1995pd}.
That is, obeying this condition is not enough to exclude
any CCB minimum to which tunnelling can occur.
It is beyond the scope of this article
to pin down the necessary and sufficient conditions for
vacuum stability with respect to the flavour-violating trilinear terms.
In any case, the stability bound is not the primary concern here.

% K physics
% B physics
%% bsg
%% B -> phi K
% LFV
%% meg
%%% mass reach of MEG exp
%% tmg, teg
% t physics
%% tVc tVu ggtc ggtu

\begin{table}
  \centering
  \begin{tabular}{c|@{\extracolsep{5ex}}ll}
    \hline
    MI &
    \multicolumn{1}{@{\hspace{-4ex}}c}{Metastability} &
    \multicolumn{1}{@{\hspace{-2.5ex}}c}{Naturalness} \\
    $\ded{12}{LR}$ & $0.026$ & $0.0011$ \\
    $\ded{13}{LR}$ & $0.026$ & $0.0010$ \\
    $\ded{23}{LR}$ & $0.026$ & $0.010$  \\
    $\deu{12}{LR}$ & $0.21 $ & $0.011$  \\
    $\deu{13}{LR}$ & $0.29 $ & $0.062$  \\
    $\deu{23}{LR}$ & $0.29 $ & $0.59$   \\
    \hline
  \end{tabular}
  \caption{Metastability and naturalness bounds for $\msq = \mgl = 1\TeV$.
    The metastability bounds on $\ded{ij}{LR}$
    are proportional to $\cos\!\beta$ and the displayed values are
    for $\tb = 10$.}
  \label{tab:naturalness}
\end{table}
There is another class of theoretical bounds that scale as the
inverse power of the sparticle masses.
They are naturalness bounds that act on half of the flavour-changing
squark mass insertions, $(\deltax{d,u}_{ij})_{LR}$ with
$i < j$ \cite{Crivellin:2008mq}.
They are based on the requirement that the supersymmetric loop corrections
do not exceed the measured values of the CKM matrix elements.
They are compared with the tunnelling limits in
Table~\ref{tab:naturalness}.
One can notice that metastability is tighter on $\deu{23}{LR}$
and that the two types of constraints are comparable on $\ded{23}{LR}$
especially for high $\tb$.
For the remaining four insertions, naturalness tends to be stronger.
This strength depends on the degree of fine-tuning
that one is willing to allow, which by contrast has nothing to do with
the metastability bounds.
Note that the slepton sector and the other half of the squark sector
$LR$ insertions are not restricted by the naturalness in renormalisation of
the lepton or the quark mixing matrix.
These insertions can be constrained by
two-loop corrections to light fermion masses but
only in a combination with another insertion \cite{Crivellin:2010gw}.
These multiple insertion bounds shall be left out
of the present comparison as
the vacuum lifetime limits a $LR$ insertion even if
it is the only non-vanishing component.

\section{Conclusions}
\label{sec:conc}

In the context of the MSSM,
upper bounds on the flavour-violating trilinear soft terms
have been obtained by demanding that the standard vacuum be long-lived.
Obviously, metastability leaves more room than stability.
As with the latter, however,
the former results in a limit that does not decouple
even if $\msf$ increases.
A distinct property of the new bounds
is that those on the 12 components $\Ade{ij}$ are nearly the same.
This is because the bounce almost follows a $D$-flat direction
and the corresponding Yukawa couplings are negligible.
% The constraints on $\Au{12,21}$ are slightly stronger since
% the other type of Higgs couples.
% The trilinears involving the stop tend to have weakest bounds
% due to the large top Yukawa coupling.
% Raising $\MH$ loosens limits on $\Ade{ij}$ but not on $\Au{ij}$ very much.
Dependence on other parameters has been reported as well
such as Higgs masses, $\tb$, and $\mu$.
The bounds are fairly stable against radiative corrections
in view of the estimated errors.

Prospects for indirect new physics discovery have been discussed
in a scenario where flavour-violating trilinear terms can be
arbitrarily large within the limits from vacuum longevity.
% Provided that one allows for arbitrary flavour-violating trilinear couplings
% within the limits from vacuum longevity,
Being sensitive to mass ranges as high as $60\TeV$ or more,
$\meg$ could be the most powerful probe of new physics.
The potential of a SFF has also been considered.
It should be able to cover up to a few TeV
through $B$ physics and LFV $\tau$ decays.
Lower $\tb$ tends to extend the territory, but it does not seem to grow
an order of magnitude beyond the reach of the LHC\@.
It should be reminded that these conclusions are based on the
assumption that the $LR$ insertions
are the only source of extra flavour violation.
As for the stop couplings,
metastability provides bounds that are stronger than any
existing experimental constraints even for rather light squarks.

\begin{acknowledgments}
  The author thanks
A.~Ali,
% Babu,
S.~Baek,
A.~Crivellin,
K.~Fujikawa,
M.~Hazumi,
N.~Kevlishvili,
% Ko, <- for discouragements, jaja
O.~Lebedev,
% Mishima,
and
H.~Nakano,
for inspirational questions and comments.
\end{acknowledgments}

%%%%% separated into an independent article
%
% \appendix*
%
% \section{Numerical computation of a bounce}
% 
% \cite{Konstandin:2006nd}
% 
% % Using a variant of K&H
% %% advantages: really computes a bounce -> answer or nothing
% %% modification: use constraint on V instead of improved potential
% %% -> starts with a profile in which the plateau has been truncated
% %%% need less lattice points -> quicker computation
% %? criticize Kusenko's method
% %% its proof is wrong
% % Initial profile (for 1D)
% %% Valley trace
% %% Do not need to know where the true vacuum is
% 
% 
% \cite{ipopt}

\end{document}